\documentclass[12pt]{article}
\usepackage{epsfig}
\usepackage{fullpage}
\begin{document}

\begin{flushright}
Report NPI \v{R}e\v{z}-TH-06/2000
\end{flushright}

\vskip 4.2cm

\begin{center}

{\Large  Z scaling, fractality, and principles of relativity
in the interactions

of hadrons and nuclei at high energies}
\vskip 1.2cm

I.Zborovsk\'y\footnote{e-mail: zborovsky@ujf.cas.cz}
\vskip 0.5cm
Nuclear Physics Institute,
Academy of Sciences of the Czech Republic,
\newline
\v{R}e\v{z}, Czech Republic

\vskip 2.0cm

Abstract
\end{center}
\vskip 0.4cm

The formation length of particles produced in the
relativistic collisions of hadrons and nuclei has relevance to
fundamental principles of physics at small interaction distances.
The relation is expressed by a $z$ scaling
observed in the differential cross sections for the inclusive
reactions at high energies.
The scaling variable reflects the length of the elementary particle
trajectory in terms of a fractal measure.
Characterizing the fractal approach, we demonstrate the relativity
principles in space with broken isotropy.
We derive relativistic transformations accounting for
asymmetry of space-time expressed by different anomalous
fractal dimensions of the interacting objects.

\vskip 2.2cm

%{\Large
%\begin{center}
%Nuclear Physics Institute,
%\end{center}
%\begin{center}
%Academy of Sciences of the Czech Republic,
%\end{center}
%\begin{center}
%250 68 že', Czech Republic
%\end{center}
%\begin{center}
%December 18th, 2000
%\end{center}
%}

\newpage
{\section{Introduction}}

Observation of particles with large transverse momenta
produced in high energy collisions of hadrons and nuclei
provides  unique information about
the  properties of quark and gluon interactions.
As follows from numerous studies in relativistic
physics (see e.g. \cite{QM93,QM96,QM99}), a common feature of the
processes is the local character of the hadron interactions.
This leads to a conclusion about
dimensionless constituents participating in the collisions.
Fact that the interaction is local manifests naturally
in a scale-invariance of the interaction cross sections.
The invariance is a special case of the `automodelity'
principle which is an expression of self-similarity
\cite{Baldin,Matveev}.
This property enables to predict and study
various phenomenological regularities reflecting the point-like
nature of the underlying interactions. A special role play the
deep inelastic processes which confirm the idea of hadrons
as composite extended objects with internal degrees of freedom.
In the collisions with nuclei the structure was revealed at
the sub-nucleon level.

The ideas were implemented into the formulation of the $z$ scaling
\cite{Z96} for description of the inclusive
particle production at high energies.
The concept of self-similarity of the interactions was
complemented by considerations about the fractal character of the
objects undergoing the collisions \cite{Z99}.
The goal of the paper is to focus on the general premisses
of the $z$ scaling in view of fundamental principles of physics
at small interaction distances. It concerns the scale dependence
of physical laws gradually emerging in various experimental and
theoretical investigations.
Such extension of physics is intrinsically linked to the
evolution of the concept of space-time.
It has been proposed \cite{Wheeler,Hawking} that the topology of
space-time becomes complicated with decreasing scales. In the extreme
case, when approaching the Plank scale, the structure of
space-time is supposed to be extremely irregular (foam-like).
The structure is characterized by explicitly
scale dependent metric potentials. Asking questions about
the metrics leads one to question the relativity.
There exists suggestions \cite{Ellis} that the
conventional special-relativistic relation between momentum and
energy might be modified reflecting `recoil' background metric
changes during the high energy interactions. This in turn could dissolve
some cutoff problems in the ultra-relativistic region.

The basic assumption tackled in the
paper is the principle of relativity which, besides motion,
applies also to the laws of scale.
There is one mathematical concept
expressing the self-similarity and complying with
the scale relativistic approach, so-called fractals \cite{Mandelbrot}.
The geometrical objects model the internal parton structure of
hadrons and nuclei revealed in their interactions at high energies.
The description of the parton interactions as expressed by
the construction of the $z$ scaling is  presented in Sec. II.
Some aspects of the fractality in view of the scaling
variable are discussed in Sec. III.
In Sec. IV., we present a formalism resulting from application
of the relativity
principles to the space with broken isotropy which is a
characteristic feature of a more fundamental concept of
fractal space-time.
One of the consequences of the approach is
a dispersion law connecting the energy and momentum of a
particle having implications to the non-standard relations of the
quantities to the particle's velocity.
In Sec. V., we apply the results to the kinematic variables
determined by the starting assumptions in the definition of the
scaling variable $z$.
As a result we obtain the relation between the coefficient expressing
the space-time anisotropy induced in the interaction
and the ratio of the anomalous (fractal) dimensions of the
interacting fractal objects.

\vskip 0.5cm
{\section{Z scaling and constituent interactions}}

We consider the high energy collisions of hadrons and nuclei as
an ensemble of individual interactions of their constituents. The
constituents can be regarded as partons in the parton model or
quarks and gluons which are building blocks in the theory
of QCD.
The presented approach is substantially based on a premise
about fractal character concerning the parton (quark and gluon)
content of the composite structures involved.
The interactions of the constituents
are local relative to the resolution which is a function of the
kinematical characteristics of the particles produced in the
collisions.
The kinematical variables characterize the resolution
in terms of the underlying processes.
They are determined in a way accounting for maximal relative
number of initial parton configurations which can lead to the
production of the observed secondaries.
In the case of the single-particle inclusive production,
the construction was presented in Ref. \cite{Z99}. It reflects
the principles of locality, self-similarity and fractality
which govern the collisions of hadrons and nuclei at high
energies.

The locality principle is widely accepted and follows
from numerous experimental and theoretical investigations.
In accordance with this principle, it has been suggested
\cite{Stavinsky} that gross features of the
single-inclusive particle distributions for the reaction
\begin{equation}
M_1+M_2 \rightarrow m_1 + X
\label{eq:r1}
\end{equation}
can be described in terms of the corresponding kinematical
characteristics of the interaction
\begin{equation}
(x_1M_1) + (x_2M_2) \rightarrow m_1 +
(x_1M_1+x_2M_2 + m_2).
\label{eq:r2}
\end{equation}
The $M_1$ and $M_2$ are masses of the colliding
hadrons (or nuclei) and $m_1$ is the mass of the inclusive
particle.
The parameter $m_2$ is used in connection with internal
conservation laws (for isospin, baryon number, and strangeness).
The $x_{1}$ and $x_{2}$ are the scale-invariant fractions of the
incoming four-momenta $P_{1}$ and $P_{2}$ of the colliding objects.
Considering the process (\ref{eq:r2}) as a local collision of the
constituents, we have exploited the coefficient
\begin{equation}
\Omega(x_1,x_2) =m^2 (1-x_1)^{\delta_1}(1-x_2)^{\delta_2}
\label{eq:r4}
\end{equation}
as a quantity proportional to the number of all initial constituent
configurations which can lead to the production of the inclusive
particle $m_1$. Here $m$ is a mass constant (its typical value
being the nucleon mass).
The coefficient connects the kinematical characteristics of the
elementary interaction with global parameters of the colliding
hadrons (or nuclei). It reflects the fractal character of the
structural objects revealed by the interaction.
The parameters $\delta_1$ and $\delta_2$ relate to their anomalous
(fractal) dimensions.

The momentum fractions $x_1$ and $x_2$ are determined in the way to
maximize the value of $\Omega$, accounting simultaneously
for the minimum recoil mass hypothesis \cite{Stavinsky}
in the constituent interaction.
The hypothesis states that the interaction is a binary collision
subjected to the condition
\begin{equation}
(x_1P_1 + x_2P_2 - q)^{2} = (x_1M_1 + x_2M_2 + m_2)^{2} ,
\label{eq:r5}
\end{equation}
where $q$  is the four-momentum of the inclusive particle
with the mass $m_1$.
The fractions resulting from these requirements were presented
in Ref. \cite{Z99}. They have the form
\begin{equation}
x_1 = \lambda_1 + \chi_1 , \ \ \ \ \ \ \
x_2 = \lambda_2 + \chi_2 ,
\label{eq:r6}
\end{equation}
where
\begin{equation}
\lambda_1 = \frac{(P_{2}q)+M_2m_2}{(P_{1}P_{2})-M_1M_2} ,\ \ \ \ \
\lambda_2 = \frac{(P_{1}q)+M_1m_2}{(P_{1}P_{2})-M_1M_2} .
\label{eq:r7}
\end{equation}
The $\chi_i$ can be expressed as follows
\begin{equation}
\chi_1 = \sqrt{\mu_1^{2}+\omega_1^{2}}-\omega_1 , \ \ \ \ \ \
\chi_2 = \sqrt{\mu_2^{2}+\omega_2^{2}}+\omega_2 ,
\label{eq:r8}
\end{equation}
where
\begin{equation}
\mu_1^{2} = \alpha\lambda^{2}\frac{(1-\lambda_1)}{(1-\lambda_2)} ,
\ \ \ \ \ \ \ \ \
\mu_2^{2} =
\frac{\lambda^{2}}{\alpha}\frac{(1-\lambda_2)}{(1-\lambda_1)} ,
\label{eq:r9}
\end{equation}
\begin{equation}
\omega_1 =
\frac{(\alpha-1)}{2}\frac{\lambda^{2}}{(1-\lambda_2)} , \ \ \ \ \
\omega_2 =
\frac{(\alpha-1)}{2\alpha}\frac{\lambda^{2}}{(1-\lambda_1)} ,
\label{eq:r10}
\end{equation}
with
\begin{equation}
\lambda = \sqrt{\lambda_1\lambda_2+\lambda_0},\ \ \ \
\lambda_0 = \frac{0.5(m_2^{2}-m_1^{2})}{(P_{1}P_{2})-M_1M_2} .
\label{eq:r11}
\end{equation}
Both $x_1$ and $x_2$ consist of two terms.
The first terms, $\lambda_i$, represent the leading parts of the
interacting
constituents and do not depend on  $\delta_1$ and $\delta_2$.
The second terms are functions of the parameter $\alpha$, where
$\alpha=\delta_2/\delta_1$
is the ratio of the corresponding anomalous
(fractal) dimensions of the objects undergoing the interaction.
It characterizes a relative state of the scale of the reference
system and is given by the type of colliding hadrons or nuclei.
The terms $\chi_i$ represent some sort of a parton `coat' of the
leading parts forming the interacting constituents.
They consist of a net of tiny partons forming a fractal structure.
The energy, momentum and mass of the parton coat depends on the
scale of the reference system characterized by the
value of $\alpha$.
According to the decomposition, the binary
subprocess (\ref{eq:r2}) can be rewritten to the symbolic form
\begin{equation}
(\lambda_1+\chi_1) + (\lambda_2+\chi_2) \rightarrow
(\lambda_1+\lambda_2)+(\chi_1+\chi_2).
\label{eq:r12}
\end{equation}
The variables $\chi_i$ correspond to the parts of the
interacting constituents which are responsible for the creation of
the recoil $(x_1M_1+x_2M_2+m_2)$.
In high energy collisions, one can, in principle,
recognize the interaction of the constituents which underlies the
inclusive particle production at high energies.
The level of the recognition is given by a resolution
$\varepsilon^{-1}$.
The resolution should depend on the extensive parameters of the
interacting system, on the available energy and should be a function of the
kinematical characteristics of the observed secondaries.
The extensive parameters in the nucleus-nucleus collisions
are the atomic numbers $A_1$ and $A_2$.
Results of our analysis on $pA$ interactions show \cite{Z99} that the
anomalous (fractal) dimensions of nuclei are
given by the relation  $\delta_A = A\delta$.
According to the additive property, the coefficient
$\alpha\equiv\delta_2/\delta_1=A_2/A_1$ is the function of the
extensive characteristics of the system but does not depend on the
structure parameter $\delta$.
Consequently, the kinematical variables $x_i$ do not depend on $\delta$.
This property enables us to define the resolution $\varepsilon^{-1}$
by means of
\begin{equation}
\varepsilon(x_1,x_2) \equiv
(1-x_1)^{A_1}(1-x_2)^{A_2} \sim \Omega^{1/\delta},
\label{eq:r12a}
\end{equation}
where $x_1$ and $x_2$ are subjected to the constrain (\ref{eq:r5}).
The $\varepsilon(x_1,x_2)$ is the relative uncertainty with
which one can single out the binary subprocess (\ref{eq:r2})
from the system of two colliding nuclei.
The uncertainty is related to the relative number of all initial
configurations of the constituents (given by $\Omega$),
interactions of which can lead to the production of the inclusive
particle $m_1$. The relative uncertainty in the determination of
the underlying subprocess can be reduced to smaller subsystems, if the
momentum fractions are not too large.  Specially, for the
interactions involving only single nucleons, one usually
introduces the momentum fractions of the interacting nuclei
expressed in units of the nucleon mass, $\bar{x}_i=A_ix_i$.
In the single nucleon interaction regime,
the relative uncertainty can be approximated as follows
\begin{equation}
\varepsilon(x_1,x_2) \equiv
(1-\bar{x}_1/A_1)^{A_1}(1-\bar{x}_2/A_2)^{A_2} \sim
(1-\bar{x}_1)(1-\bar{x}_2) .
\label{eq:r12b}
\end{equation}
On the other hand, the factorization does not apply for
the processes in which $\bar{x}_i>1$.
They are known as cumulative processes \cite{Baldin,Stavinsky,Leksin}
and correspond to the joining of partons from different
nucleons of nuclei.
This region is interesting from the point of view of
fractality at small scales.
Here we focus on small interaction distances under the condition
that simultaneously large amount of energy is deposited in it.
In the single-particle inclusive reactions, the cumulative process
corresponds to larger spatial resolutions determined by the
dimensions of the interaction region of the constituent collision.
The resolution increases with the momentum transfer in the underlying
subprocesses and is influenced by the types of the colliding
objects. We discuss this connection in Appendix A in more
detail.

The second basic ingredient of the $z$ scaling scheme is
the self-similarity principle.
In high energy physics, the principle is reflected by the
property that the underlying production processes are similar.
It results in dropping of certain quantities or parameters out
of the physical picture of the interactions.
The particle production from self-similar processes can be described
in terms of independent variables.
Construction of the variables depends on the type of the
inclusive reaction.
For single-particle inclusive production at high energies,
the scaling function \cite{Z99}
\begin{equation}
H(z) \equiv \frac{\psi(z)}{2\pi z} =
\frac{s_{tot}}{\rho_{tot}\sigma_{inel}}
\left(
\frac{z}{\lambda_2}\frac{\partial z}{\partial\lambda_1}
+\frac{z}{\lambda_1}\frac{\partial z}{\partial\lambda_2}
\right)^{-1}
E\frac{d^3\sigma}{d{q}^3}
\label{eq:r13}
\end{equation}
was studied in dependence on the single variable $z$.
Here $s_{tot}$ is a square of the total center-of-mass energy and
$\sigma_{inel}$ is the inelastic cross section of the
objects $M_1$ and $M_2$.
The relation connects the inclusive differential
cross section and the multiplicity density
$\rho_{tot}(s_{tot},\eta)\equiv d\langle N\rangle/d\eta$
of the corresponding pseudorapidity distribution with the
scaling function.
The function $\psi(z)$ is interpreted as the probability
density to form particles characterized by the value of $z$.
The question is to define such $z$ that could reflect
general pattern of the particle production mechanism.
The variable was chosen accounting for the locality,
self-similarity and fractality as common principles of the
particle production at high energies \cite{Z99}.
It represents the fractal measure attributed to the
inclusive particle and is proportional to its formation length.
The measure consists of a finite part $z_0$ and a divergent factor
depending on the resolution $\varepsilon^{-1}$.
It is defined in the form
\begin{equation}
z = z_0 \ \varepsilon^{-\delta} ,
\label{eq:r14}
\end{equation}
where
\begin{equation}
z_0 = \frac{\hat{s}^{1/2}_{\bot kin}}{m^2\rho(s)} , \ \ \ \ \ \ \ \ \ \
\varepsilon(x_1,x_2) \equiv
(1-x_1)^{A_1}(1-x_2)^{A_2} .
\label{eq:r15}
\end{equation}
The finite part $z_0$ is characterized by the
transverse kinetic energy
\begin{equation}
\hat{s}^{1/2}_{\bot kin} =
\hat{s}^{1/2}_{\lambda} + \hat{s}^{1/2}_{\chi}
-m_1-(M_1x_1+M_2x_2+m_2)
\label{eq:r16}
\end{equation}
of the subprocess (\ref{eq:r2}). It includes the terms
\begin{equation}
\hat{s}^{1/2}_{\lambda} =\sqrt{(\lambda_1P_1+\lambda_2P_2)^2},
\ \ \ \ \ \
\hat{s}^{1/2}_{\chi} =\sqrt{(\chi_1P_1+\chi_2P_2)^2},
\label{eq:r17}
\end{equation}
representing the transverse energy of the inclusive particle and
its recoil, respectively.
The factor $\rho(s) \equiv d\langle n \rangle/d\eta|_{\eta=0}$
is the average multiplicity density of the charged
particles produced in the
central region of the corresponding nucleon-nucleon interaction.
The multiplicity density $\rho(s)$ depends on the total
center-of-mass energy and includes the dynamical ingredient of
the scaling.
The divergent part of the variable $z$ is given as a $\delta$
power of the resolution $\varepsilon^{-1}$.
The parameter $\delta$ represents anomalous (fractal)
dimension of the trajectories followed by the inclusive particles
produced in the high energy collisions of hadrons and nuclei.
We consider the anomalous dimension $\delta$ to be a characteristic
reflecting the intimate structure of particle motion at small
scales and thus related to the fractal properties of space-time.

The comparison with experimental data shows that the
$z$ scaling represents regularity valid in nucleon-nucleon and
proton-nucleus collisions in a wide range of the
center-of-mass energy,
the detection angle $\theta$ and the atomic number $A$.
All parameters in the $z$ scaling scheme are given in terms
of the measurable quantities except one, which is the
parameter $\delta$. Restriction on this
parameter is given by  experimental data.
We stress that energy and angular independence of the $z$ scaling
can be achieved simultaneously by the same value of $\delta$.
We have obtained $\delta\sim 0.8$ \cite{Z99} for the inclusive production
of charged particles (pions). The A-universality of the $z$ scaling
was demonstrated with the same value as well.
The simultaneous energy and angular independence of $z$ scaling
for the inclusive productions of jets implies \cite{Tokarev}
$\delta\sim 1$.
It should be emphasized that the production of jets underlie the hard
processes with large values of $z$. In this region the scaling
function  manifest a clear power law $\psi(z)\sim z^{-\beta}$.
Similar behaviour is seen for the inclusive production of single
hadrons \cite{Z99} in the tail of the spectrum.
The power law of the scaling function stresses the fractal attributes
of the processes  which are preferable to study mainly in the
region of large $z$.
In contrast to this, the scaling function has different shape for
small values of $z$. It corresponds to medium and low
transverse momenta $q_{\bot}$. The underlying parton interactions
have character of the soft processes here. The soft interactions
are dominant mainly in both fragmentation regions.
In this sense, the soft and hard regime of the particle production
is distinguished by different behaviour of the scaling function.
The quantitative values of the scaling function should be reproduced
using a theory of hadronic interactions, which we thrust is QCD.
However difficulties in applying the QCD methods in the non-perturbative
regime and the lack of fundamental understanding of hadronization
even in an perturbative approximation, limits us to the phenomenology.
In our description of the inclusive reactions, we aim at grasping
main principles that influence the particle production at small
scales.
We understand the existence of the $z$ scaling itself as confirmation
of the hadron interaction locality, self-similarity, and fractality
which possess an universal character.

\vskip 0.5cm
{\section{Fractality at small distances}}

Besides the self-similarity principle, the construction of the
$z$ scaling regards also the fractality  of the hadron interactions.
The fractal properties are manifested especially at small interaction
distances.
In this region, physics becomes scale dependent and possesses
its typical property which is the divergence of elementary
quantities such as the self-energy, charges and so on.
In the high energy collisions, one of the relevant physical
quantities is the formation length of the produced particles
divergent at small scales.
According to the $z$ scaling hypothesis, the formation length is
proportional to the scaling variable $z$ and the production cross
sections depend on it in an universal, energy independent way
\cite{Tokarev,T99,Z94}.

Universality plays the crucial role.
There exist suggestions \cite{Nottale} that universal properties
of matter are attributed to the structure of space-time itself.
In the theory of
relativity both special and general, this concerns the Lorenz
transformation and the curvature of the trajectories reflecting
fundamental principles of physics. Free particles are moving
along smooth geodesical lines, characteristic for the classical
(curved) space-time.
The situation becomes different at scales typical for the quantum
world of the elementary particles. The common property is
the unpredictability of motion at small distances. In this region
the particles follow irregular trajectories which become
non-differentiable.
The geometry of the lines of motion can be
attributed to the properties of fractals which are extremely
irregular objects fragmented at all scales. As an example one can
mention the quantum-mechanical path of
a particle in the sense of Feynman trajectories \cite{Feynman}.

One of the main characteristics of fractals is the divergence of
their measures in  terms of the increasing resolution.
We illustrate this property  by the von Koch curve \cite{Koch}
which is characterized by the fractal measure
\begin{equation}
z_{\varepsilon} = z_0 \cdot \varepsilon^{D_T-D}
\label{eq:r18}
\end{equation}
representing its length.
The length of the curve is a function of the resolution
$\varepsilon^{-1}$ (see Fig.1.).
Relation (\ref{eq:r18}) is typical for various fractals
and states how the fractal measure depends on the resolution.
Fractal objects are characterized by the topological dimensions
$D_T$  which are integer quantities and by the fractal dimensions
$D$ acquiring generally non-integer values.
The anomalous dimension of a fractal
\begin{equation}
\delta \equiv D - D_T > 0
\label{eq:r19}
\end{equation}
is positive. It is equivalent to power law divergence of the measure
$z_{\varepsilon}$ with the increasing resolution.
Von Koch curve illuminates the process of
`fractalization'. The curve has the length
\begin{equation}
z_n = z_0 (p/q)^{n}, \ \ \ \ p=4, \ \ q=3
\label{eq:r20}
\end{equation}
in the n-th approximation. It is composed of $p^{n}$ segments,
each of the length $z_0q^{-n}$.
The measure can be rewritten to the form
\begin{equation}
z_n = z_0 (q^{-n})^{1-\ln p /\ln q} ,
\label{eq:r20a}
\end{equation}
which gives the relation (\ref{eq:r18}) with $D_T=1$,
$D=\ln p /\ln q $, and $\varepsilon = q^{-n}$.
The anomalous dimension of the von Koch curve $\delta$ is
positive and thus, with increasing resolution, its length tends to
infinity.

These concepts find application in the world of physics at small
distances.
The fractal character in the initial state reflects the
parton (quark and gluon) composition of hadrons and nuclei
and reveals itself with a larger resolution at high energies.
According to this picture, the collisions of the constituents take
place on the fractal background of the interacting objects.
The essential assumption concerning the interpretation of the
ideas represented by the $z$ scaling hypothesis is expressed in
the following statement:
{\it Presence of the interacting fractal objects deforms the
structure of surrounding space at small distances.
As a consequence, space-time becomes locally non-differentiable,
fractal with geodesical lines acquiring an extremely
irregular scale-dependent shape.}
The notion of fractal space-time was used in Refs. \cite{Nottale,Ord}
and its properties have been studied by others \cite{Pissondes}.
Distortion of the surrounding space-time in the interactions with
recoil induce a non-trivial off-diagonal terms in the metric
changing the relations between energy and momentum \cite{Ellis}.
The issues concern modelling of various aspects of quantum
fluctuations which influence particle production and their
interactions at small scales.

The secondary partons, produced in fractal space-time follow
erratic and scale-dependent geodesics starting from the regions they
have been created.
During the initial phase of the motion, one can imagine the
trajectories as fractal curves similar to the von Koch curve
depicted in Fig.1.
Formation of a particle from the bare parton
realizes along the trajectory characterized by its length.
The produced parton, ancestor of the secondary particle, interacts
with the vacuum or the surrounding matter field acquiring
simultaneously some type of the parton `coat'. The number of the
enveloping partons forming the coat results in an effective increase
of the ancestor's dimensions simultaneously changing its mass.
During the particle formation process, the path of the leading parton
becomes gradually smoother in comparison to that following
immediately the instant of the constituent interaction.
In the final stage of the process, the relative length of the
particle's path is negligible with respect to its value in the very
beginning of the motion.
It is a consequence of the transition from small scales,
characterized by an extremely irregular fractal-like character
of the trajectory, to scales larger than, say, the
corresponding de Broglie length.
Thus, in the collisions of the structural objects such as
hadrons or nuclei, the length of the trajectory of a produced
particle can increase infinitely with the decreasing scale at which
the particle was created.

Let us look at the problem in view of the models
describing the interaction of the constituents as a parton-parton
collision with subsequent formation of a string stretched
between the two parts in the final state of the process
(\ref{eq:r2}).
Following the same geodesical trajectory, the string is
a fractal object with the scale dependent properties.
In this picture, the value of $\Omega$ (\ref{eq:r4}) can be
considered as a quantity reflecting the tension of the string.
The string tension coefficient is
characterized by the density of the straight-like segments
along its length. For all resolutions, the product of the coefficient
and the length of the string $z_{\varepsilon}$ represents the finite
quantity
\begin{equation}
\hat{s}^{1/2}_{h} = \Omega \cdot z_{\varepsilon} ,
\label{eq:r20b}
\end{equation}
which is the energy of the string. The form of the tension
coefficient, as presented by Eq. (\ref{eq:r4}), accounts for the
shape of the string deformed in consequence of the fractal structure
of space-time.
With the increasing resolution, the string is more and more
fragmented and the deformations result in the diminishing of
its tension. Physically, the deformations of the string can be caused
by fluctuations of the
QCD vacuum disturbed in presence of the interacting fractal objects
at small distances. The string tension
\begin{equation}
\Omega \sim \varepsilon^{\delta}
\label{eq:r20c}
\end{equation}
is a function of the resolution $\varepsilon^{-1}$
and is characterized by the anomalous (fractal) dimension
$\delta$ of the string.
The resolution corresponds to a characteristic size of the space-time
region of the interaction and depends on the momentum fractions
carried by the interacting constituents.
In the determination of the fractions, we consider an optimization
method
dealing with the fractal trajectories of particles at any scale.
The optimal trajectory is defined by the condition that, for
the underlying collision of the constituents, the tension of the
fractal-like string, stretched out by the produced parton, should by
maximal.
This is equivalent to the extremum of the expression (\ref{eq:r4})
under the condition (\ref{eq:r5}).
The maximal tension of the string is thus given by the minimal value
of the resolution and corresponds to the geodesics which are in a
sense `optimal' curves.

The energy of the string connecting the two objects
in the final state of the process (\ref{eq:r2}) is given by
the energy of the colliding constituents.
The string evolves further and splits into pieces.
The resultant number of the string pieces is proportional
to the number or density of the final hadrons measured in experiment.
As known from various experimental and theoretical studies
concerning the multiple production, the produced multiplicity is
proportional to the excitation of transverse degrees of freedom.
Therefore, the string transverse energy is a measure of multiplicity.
Such ideas allow us to interpret the ratio
\begin{equation}
\hat{s}^{1/2}_{h} \equiv \hat{s}^{1/2}_{\bot kin}/\rho(s)
\label{eq:r20d}
\end{equation}
as a quantity proportional to the energy of the string piece,
which does not split already, but during the formation process
converts into the observed secondary.
The string splitting is self-similar in the sense that the leading
piece of the string forgets the string history and its formation
does not depend on the number and behaviour of other pieces.
We write the energy of the single piece of the string in the form
(\ref{eq:r20b}).
This is equivalent to the definition of the variable
$z\equiv z_{\varepsilon}$ as a fractal measure proportional to the
length of the string piece, or the formation length, on which the
inclusive particle is formed.
The corresponding scaling function $H(z)$ reflects the evolution of
the formation process of the inclusive particle along its
fractal-like trajectory of the length $z$.
We add here that there exists also complementary interpretation
of the factor $\Omega$.
According to the ideas presented in Ref. \cite{Z99},
$\Omega$ reflects the relative number of
all initial configurations containing the constituents
which carry the momentum fractions $x_1$ and $x_2$. The number of
the configurations in one colliding object is given by the
power law characteristic for fractals.
In fractal dynamics the resolution $\varepsilon^{-1}$ is given by
the maximal number of the initial configurations which can lead to
the production of the particle $m_1$.

Generally, the fractal approach to the high energy
collisions of hadrons and nuclei needs more profound understanding.
It concerns the deformation of
space-time at small scales and attributes additional meaning
to the physical quantities such as the momentum, mass, energy or
velocity. They may be defined from parameters of the fractal
objects in terms of the fractal geometry \cite {Nottale}.
This includes extension of the relativity principles to the
relativity of scales as well as to the more comprehensive
scale-motion relativistic concepts.

\vskip 0.5cm
{\section {Break down of the reflection invariance, the way towards
scale-motion relativity}}

General solution to the theory of the special relativity is the
Lorenz transformation. As demonstrated by Nottale, it can be
obtained under minimal number of three successive constraints.
They are (i) homogeneity of space-time translated as the linearity
of the transformation, (ii) the group structure defined
by the internal composition law and (iii) isotropy of
space-time expressed as the reflection invariance.
Let us consider the relativistic boost along the x-axis.
The transformation concerns the variables $x$ and $t$
which refer not only to the coordinate and time, but also to any
quantities having the mathematical properties considered.
Without any loss of generality, the linearity of the
transformation can be expressed in the form \cite{Lalan}
\begin{equation}
x'  = \gamma (u)[x-ut] ,
\label{eq:r21}
\end{equation}
\begin{equation}
t'  = \gamma (u)[A(u)t-B(u)x] ,
\label{eq:r22}
\end{equation}
where $\gamma$, $A$, and $B$ are functions of a parameter $u$.
The parameter represents usual velocity (in units of the velocity
of light c) in the motion relativity
or the `scale velocity' used, e.g., in the concept of the scale
relativity concerning fractal dimensions and fractal
measures \cite{Nottale}.
Let us compose the transformation with the successive one
\begin{equation}
x''  = \gamma (v')[x'-v't'] ,
\label{eq:r23}
\end{equation}
\begin{equation}
t''  = \gamma (v')[A(v')t'-B(v')x'] .
\label{eq:r24}
\end{equation}
The result can be written in the form
\begin{equation}
x''  = \gamma(u)\gamma(v')[1+B(u)v']
\left[x-\frac{u+A(u)v'}{1+B(u)v'}t\right] ,
\label{eq:r25}
\end{equation}
\begin{equation}
t''  = \gamma(u)\gamma(v')[A(u)A(v')+B(v')u]
\left[t-\frac{A(v')B(u)+B(v')}{A(u)A(v')+B(v')u}x\right] .
\label{eq:r26}
\end{equation}
The principle of relativity is expressed by the constraint (ii)
which is the group structure of the transformations. The
condition tells us that Eqs. (\ref{eq:r25}) and (\ref{eq:r26}) keep
the same form as the initial ones in terms of the composed
velocity
\begin{equation}
v  = \frac{u+A(u)v'}{1+B(u)v'} .
\label{eq:r27}
\end{equation}
The requirement can be satisfied under the following
conditions
\begin{equation}
\gamma(v)  = \gamma(u)\gamma(v')[1+B(u)v'] ,
\label{eq:r28}
\end{equation}
\begin{equation}
\gamma(v)A(v)  = \gamma(u)\gamma(v')[A(u)A(v')+B(v')u] ,
\label{eq:r29}
\end{equation}
\begin{equation}
\frac{B(v)}{A(v)} = \frac{A(v')B(u)+B(v')}{A(u)A(v')+B(v')u} .
\label{eq:r30}
\end{equation}
The third constraint, the isotropy of space-time, results in
the requirement that the change of orientations of the variable axis are
indistinguishable, provided $u'=-u$. This leads to the parity
relations $\gamma(-u)=\gamma(u)$, $A(-u)=A(u)$, and
$B(-u)=-B(u)$. The relations are sufficient for the derivation
of the Lorenz transformation.
The theory of relativity tells us that the velocity of a
physical object can not exceed the value
of $c$, the velocity of light in the vacuum.
The expression of this statement is the Lorenz transformation which
yields the limitation of any velocity.
As concerns the relativity principles including the scale degrees
of freedom, we need to distinguish two approaches with respect to the
concept of motion.

The first one, the scale relativity, is based on the premise that
the relation between the energy or momentum of a particle and
its velocity is inoperative in fractal space-time.
Instead of the motion velocity, the quantities become functions of
the `velocity of scales'. The `velocity' depends on the
fractal characteristics and does not represent any real motion.
The special scale relativity concerns the description of physical
events with respect to the self-similar scale structures which are
fractals of various fractal dimensions. The corresponding Lorenz-type scale
transformations relate the physical quantities expressed in terms of
one fractal structure with the quantities given with respect to the
other one. Single fractal structures have different anomalous fractal
dimensions and play analogous role as the inertial systems in
the motion relativity.

The second approach, the scale-motion relativity, preserves
relations of the energy and momentum to the velocity.
This applies also to small scales where we assume space-time
to possess an intrinsic (fractal-like) structure.
Our complete change of view of a particle
in the corresponding fractal space-time concerns the
divergence of the fractal measure representing the length of
the particle's trajectory.
The degree of revelation of the structures depends on the
resolution. For any given resolution $\varepsilon^{-1}$,
the non-differentiable fractal space-time $F$ can be approximated
by a Riemann space $R_{\varepsilon}$ defined within
a differentiable geometry.
As pointed in Ref.\cite{Nottale},
the family of the Riemann spaces is characterized by metric tensors
curvatures of which are expected to fluctuate in a chaotic way.
The fluctuations increase with the decreasing scale.
For the high resolutions $\varepsilon^{-1}$ the approximations to
the fractal measure tend to infinity.
The corresponding length $z_{\varepsilon}$ of the particle trajectory
can be arbitrary large.
The propagation of a physical signal along such a trajectory requires
the velocities exceeding the value of $c$, the speed
of light in symmetric space-time.
Therefore, the application of the principles of relativity to
space-time with fractal properties should be treated carefully.
The significant characteristics of the fractal spaces is their fluctuating and
irregular  nature. The corresponding geodesical lines are extremely
unpredictable and fragmented at any scale. As a consequence, the isotropy of
space-time is clearly broken. This is connected with
the breaking of the reflection invariance at the infinitesimal
level \cite{Pissondes}.
The application of the ideas to space-time at small scales
leeds us to leave out the constraint (iii) of the
reflection invariance, when considering the
relativistic transformations (\ref{eq:r21}) and (\ref{eq:r22}).
In that case the unknown functions $\gamma$, $A$, and $B$
do not obey the parity relations resulting from the
isotropy requirement.
Let us combine Eqs. (\ref{eq:r27}), (\ref{eq:r28}), and
(\ref{eq:r29}) into the expression
\begin{equation}
A\left( \frac{u+A(u)v'}{1+B(u)v'}\right) =
\frac{A(u)A(v')+B(v')u}{1+B(u)v'} .
\label{eq:r31}
\end{equation}
Its solution has the form
\begin{equation}
A(u) = 1 + 2au ,
\label{eq:r32}
\end{equation}
provided $B(u)v' = B(v')u$.
The condition gives the function $B(u) = u$ with the
normalization constant $c$ included already in the
definition of the variable $u$.
The solution satisfies Eq. (\ref{eq:r30}) as well.
The violation of the space-time reflection invariance is
expressed by a non-zero value of $a$.
In terms of the parameter $a$, the composed velocity (\ref{eq:r27})
can be written as follows
\begin{equation}
v  = \frac{v'+u+2auv'}{1+uv'} .
\label{eq:r33}
\end{equation}
Using this relation, Eq. (\ref{eq:r28}) becomes
\begin{equation}
\gamma\left(\frac{v'+u+2auv'}{1+uv'}\right)
= \gamma(u)\gamma(v')(1+uv') .
\label{eq:r34}
\end{equation}
Its solution, which for $a=0$ is given by the standard $\gamma$
factor, has the form
\begin{equation}
\gamma(u) = \frac{1}{\sqrt{1+2au-u^{2}}} .
\label{eq:r35}
\end{equation}

\vskip 0.5cm
\subsection{Space-time asymmetry in 3+1 dimensions}

Let us describe a point $P$ in two Cartesian reference
systems $S$ and $S'$.
We assume that the systems are oriented parallel to each other and that
$S'$ is moving relative to $S$ with the velocity $u$ in the
direction of the positive $x$-axis. We suppose that the
asymmetry expressed by the parameter $a$ is parallel to the
velocity $u$.
The relativistic transformations of the coordinates
and time are given by
\begin{equation}
x_1' = \gamma(u) \left[x_1-ut\right] , \ \ \ \ \
x_i' = x_i, \ \ \ \ i=2,3,
\label{eq:r36}
\end{equation}
\begin{equation}
t' = \gamma(u)
\left[\left(1+2au\right)t-ux_1\right] .
\label{eq:r37}
\end{equation}
The inverse relations
\begin{equation}
x_1 = \gamma(u) \left[(1+2au)x_1'+ut'\right] ,
\ \ \ \ x_i = x_i', \ \ \ \ i=2,3,
\label{eq:r38}
\end{equation}
\begin{equation}
t = \gamma(u) \left[t'+ux_1'\right]
\label{eq:r39}
\end{equation}
are obtained as the solution of Eqs. (\ref{eq:r36}) and (\ref{eq:r37})
with respect to the unprimed variables.
They can be also derived from the equations by
the interchange $\vec{x}\leftrightarrow \vec{x}'$,
$t\leftrightarrow t'$, $u\leftrightarrow u'$, and by the relation
\begin{equation}
u' = -\frac{u}{1+2au} .
\label{eq:r40}
\end{equation}
This formula connects the velocity $u'$ of the system $S$ in the
$S'$ frame with the velocity $u$ of the system $S'$ in the $S$
reference system.
Because of the asymmetry parameter $a$, the magnitudes of the two
velocities are not equal.
For the vanishing value of $a$, the transformations turn into the usual
relativistic transformations of the coordinates and time.
One can show by the direct calculation that the invariant
of the transformations has the form
\begin{equation}
t^{2}-x^{2}_1+2tax_1 .
\label{eq:r41}
\end{equation}
If the space-time anisotropy $\vec{a}$ acquires an arbitrary direction,
we write the invariant in the more general form
\begin{equation}
\hat{a}_{\mu\nu}x^{\mu}x^{\nu} =
t^{2}-\vec{x}^{2}+
2t\vec{a}\!\cdot\!\vec{x} - (\vec{a}\!\times\!\vec{x})^2
\equiv \tau^2 .
\label{eq:r42}
\end{equation}
Besides the diagonal part, it has extra
terms given by a non-zero values of the vector $\vec{a}$.
Similar extra terms of the relativistic invariant were considered in
Ref. \cite{Pissondes} and associated with breaking of the reflection
invariance assumed at the infinitesimal level.
In Ref. \cite{Ellis}, the off-diagonal
terms in the metric are connected with distortion of
space-time by the recoil particle in the interaction.
Using the four dimensional notation,
the invariant (\ref{eq:r42}) corresponds to the metric
\begin{equation}
\hat{a}_{\mu\nu} =
\left(
\begin{array}{cc}
-d_{ij}  & a_i \\
a_j & 1 \\
\end{array}
\right), \ \ \ \ \ \ \ \
d_{ij}=(1+a^2)\delta_{ij}-a_ia_j.
\label{eq:r43}
\end{equation}
Here the indices $i$ and $j$ numerate the first three rows and
columns of the matrix $\hat{a}$, respectively.
The $\delta_{ij}$ is the Kronecker's symbol.
Next we present the explicit form for the relativistic
transformations of the coordinates and time in the considered case.
They must be linear and homogeneous, preserving
the invariant (\ref{eq:r42}).
The transformations have to possess an internal group structure
required by the principle of relativity.
We denote the parameter of the group by the symbol $\vec{u}$.
The parameter is the velocity
of the system $S'$ with respect to the $S$ reference frame.
In connection with the transformation formulae, it is convenient
to introduce the notations
\begin{equation}
\gamma = \frac{1}
{\sqrt{(1+\vec{a}\!\cdot\!\vec{u})^{2}-
(1+a^{2})u^{2}}}
\label{eq:r44}
\end{equation}
and
\begin{equation}
g = \frac{(1+\vec{a}\!\cdot\!\vec{u})\gamma-1}{u^{2}} .
\label{eq:r45}
\end{equation}
Here $a^{2}=\vec{a}\!\cdot\!\vec{a}$ and
$u^{2}=\vec{u}\!\cdot\!\vec{u}$.
Let us define the following combinations of $\gamma$ and $g$,
\begin{equation}
\gamma_{\pm} = g u^{2}\pm \gamma\vec{a}\!\cdot\!\vec{u} ,
\ \ \ \ \ \ \ \
g_{\pm} =
\gamma (1+a^{2}) \pm
g \vec{a}\!\cdot\!\vec{u} .
\label{eq:r46}
\end{equation}
Now we are ready to consider the relations
\begin{equation}
\vec{x}' = \vec{x} -
\vec{u}\left[\gamma(t+\vec{a}\!\cdot\!\vec{x}) -
g\vec{u}\!\cdot\!\vec{x}\right] ,
\label{eq:r47}
\end{equation}
\begin{equation}
t' = t +
\left[\gamma_{+}(t+\vec{a}\!\cdot\!\vec{x}) -
g_{+}\vec{u}\!\cdot\!\vec{x}\right] .
\label{eq:r48}
\end{equation}
They generalize the special transformations (\ref{eq:r36})
and (\ref{eq:r37}) which are recovered by
$\vec{u} = (u,0,0)$ and $\vec{a} = (a,0,0)$.
The inverse relations can be obtained
by the interchange $\vec{x} \leftrightarrow \vec{x}'$,
$t \leftrightarrow t'$, $\vec{u} \leftrightarrow \vec{u}'$,
and by the substitution
\begin{equation}
\vec{u}' = -\frac{\vec{u}}{1+2\vec{a}\!\cdot\!\vec{u}} .
\label{eq:r50}
\end{equation}
According to the substitution, there exist the symmetry properties
\begin{equation}
\gamma(\vec{u}') =
(1+2\vec{a}\!\cdot\!\vec{u})\gamma(\vec{u}) , \ \ \ \ \
\gamma_{\pm}(\vec{u}') = \gamma_{\mp}(\vec{u}) ,
\label{eq:r51}
\end{equation}
\begin{equation}
g(\vec{u}') =
(1+2\vec{a}\!\cdot\!\vec{u})^{2}g(\vec{u}) , \ \ \ \ \ \
g_{\pm}(\vec{u}') =
(1+2\vec{a}\!\cdot\!\vec{u})g_{\mp}(\vec{u}) .
\label{eq:r52}
\end{equation}
Exploiting the properties, the inverse relations
\begin{equation}
\vec{x} = \vec{x}' +
\vec{u}\left[\gamma(t'+\vec{a}\!\cdot\!\vec{x}') +
g\vec{u}\!\cdot\!\vec{x}'\right] ,
\label{eq:r53}
\end{equation}
\begin{equation}
t = t' +
\left[\gamma_{-}(t'+\vec{a}\!\cdot\!\vec{x}') +
g_{-}\vec{u}\!\cdot\!\vec{x}'\right]
\label{eq:r54}
\end{equation}
with respect to Eqs. (\ref{eq:r47}) and (\ref{eq:r48})
follow immediately.
We express the relativistic transformations
in a more compact form
\begin{equation}
x' = D(\vec{u})x ,
\label{eq:r55}
\end{equation}
where
\begin{equation}
D(\vec{u}) =
\left(
\begin{array}{cc}
\delta_{ij}\!+\!g u_iu_j\!-\!\gamma u_ia_j &
-\gamma u_i \\
-g_{+}u_j\!+\!\gamma_{+}a_j & 1\!+\!\gamma_{+} \\
\end{array}
\right) .
\label{eq:r56}
\end{equation}
The inverse matrix reads
\begin{equation}
D(\vec{u}') = D^{-1}(\vec{u}) =
\left(
\begin{array}{cc}
\delta_{ij}\!+\!g u_iu_j\!+\!\gamma u_ia_j &
+\gamma u_i \\
+g_{-}u_j\!+\!\gamma_{-}a_j & 1\!+\!\gamma_{-} \\
\end{array}
\right) .
\label{eq:r57}
\end{equation}
The transformation matrices can be decomposed into the product
\begin{equation}
D(\vec{u}) =
A^{-1}_x(\vec{a}) \Lambda(\vec{\beta}) A_x(\vec{a}) .
\label{eq:r58}
\end{equation}
Here
\begin{equation}
A_x (\vec{a}) =
\left(
\begin{array}{cc}
\sqrt{1+a^2}\delta_{ij} &  0 \\
a_j &  1 \\
\end{array}
\right)
\label{eq:r59}
\end{equation}
and
\begin{equation}
\Lambda (\vec{\beta}) =
\left(
\begin{array}{cc}
\delta_{ij}\!+\!g_{0} \beta_i\beta_j &
-\gamma_{0}\beta_i  \\
-\gamma_{0}\beta_j  &  \gamma_{0}\\
\end{array}
\right)  ,
\label{eq:r60}
\end{equation}
with
\begin{equation}
\gamma_{0} = \frac{1}{\sqrt{1\!-\!\beta^{2}}} , \ \ \ \ \ \ \ \ \
g_{0} = \frac{\gamma_{0}-1}{\beta^{2}}.
\label{eq:r61}
\end{equation}
The matrix $\Lambda$ depends on the vector
\begin{equation}
\vec{\beta} \equiv \vec{\beta}_{\vec{u}} =
\sqrt{1\!+\!a^2}\frac{\vec{u}}{1+\vec{a}\!\cdot\!\vec{u}} .
\label{eq:r62}
\end{equation}
Let us notice that the interchange
$\vec{u}\leftrightarrow\vec{u}'$ is equivalent to the symmetry
$\vec{\beta}\leftrightarrow -\vec{\beta}$.
The relativistic transformations (\ref{eq:r55}) preserve the
invariant (\ref{eq:r42}). This follows from the relation
\begin{equation}
D^{\dag}(\vec{u}) \hat{a} D(\vec{u}) = \hat{a} =
A^{\dag}_x\eta A_x ,
\label{eq:r63}
\end{equation}
where $\eta$ stands for the diagonal matrix
$\eta$=diag(-1,-1,-1,+1).

The transformations comply the principle of relativity.
Mathematically it is expressed by their group properties.
Let $D(\vec{u})$ and $D(\vec{v}')$ be two
successive relativistic transformations represented by the
matrices (\ref{eq:r56}).
The composition of the  transformations has the form
\begin{equation}
\Omega_x(\vec{\phi}) D(\vec{v})  =
D(\vec{v}') D(\vec{u}) ,
\label{eq:r64}
\end{equation}
provided
\begin{equation}
\vec{v} = \frac{\vec{v}' +
\vec{u}\left[\gamma(1+\vec{a}\!\cdot\!\vec{v}')
+g\vec{u}\!\cdot\!\vec{v}'\right]  }
{1+\gamma_{-}(1+\vec{a}\!\cdot\!\vec{v}')
+g_{-}\vec{u}\!\cdot\!\vec{v}'} .
\label{eq:r65}
\end{equation}
One can obtain the above relations by exploiting the
decomposition (\ref{eq:r58}) and
using the structure of the Lorenz group expressed by the
formula
\begin{equation}
R(\vec{\phi}) \Lambda(\vec{\beta}_v) =
\Lambda(\vec{\beta}_{v'}) \Lambda(\vec{\beta}_{u}) .
\label{eq:r66}
\end{equation}
The matrix
\begin{equation}
R(\vec{\phi}) =
\left(
\begin{array}{cc}
R^{\varphi}_{ij} & 0 \\
0 & 1  \\
\end{array}
\right) , \ \ \ \ \
\vec{\phi} = \vec{v}'\times\vec{u}
\label{eq:r67}
\end{equation}
describes the Thomas precession \cite{Thomas} around the vector
$\vec{\phi}$ known in the theory of relativity.
The angle of the precession $\varphi$ depends on the vectors
$\vec{\beta}_{v'}$ and $\vec{\beta}_u$.
It remains to identify
\begin{equation}
\Omega_x(\vec{\phi}) = A^{-1}_x R(\vec{\phi}) A_x
\label{eq:r68}
\end{equation}
and we get Eq. (\ref{eq:r64}).
The relativistic transformation of the coordinates and
time with rotation of the coordinate axes has the structure
\begin{equation}
D(\vec{u}) \Omega_x(\vec{\phi}) =
\left(
\begin{array}{cc}
R^{\varphi}_{ij}\!+\!g u_iu_rR^{\varphi}_{rj}
\!-\!\gamma u_ia_j &
-\gamma u_j  \\
-g_{+}u_rR^{\varphi}_{rj}\!-\!a_rR^{\varphi}_{rj}\!+\!
(1\!+\!\gamma_{+}) a_j
 & 1\!+\!\gamma_{+} \\
\end{array}
\right) ,
\label{eq:r69}
\end{equation}
provided the asymmetry of space-time is expressed by the
vector $\vec{a}$.
As concerns Eq. (\ref{eq:r65}), it can be obtained from the usual
relativistic composition of the factors $\vec{\beta}$ given by
Eq. (\ref{eq:r62}).
The inverse relation
\begin{equation}
\vec{v}' = \frac{\vec{v} -
\vec{u}\left[
\gamma(1+\vec{a}\!\cdot\!\vec{v})
-g\vec{u}\!\cdot\!\vec{v}\right]  }
{1+\gamma_{+}(1+\vec{a}\!\cdot\!\vec{v})-
g_{+}\vec{u}\!\cdot\!\vec{v}}
\label{eq:r70}
\end{equation}
corresponds to the composition of the transformations in the
following form
\begin{equation}
\Omega_x(-\vec{\phi}) D(\vec{v}') =
D(\vec{v}) D^{-1}(\vec{u}) .
\label{eq:r71}
\end{equation}
When using  Eqs. (\ref{eq:r65}) and (\ref{eq:r70}),
we get
\begin{equation}
1+\vec{a}\!\cdot\!\vec{v} = \gamma
\frac{(1+\vec{a}\!\cdot\!\vec{u})
(1+\vec{a}\!\cdot\!\vec{v}')+
(1+a^{2})\vec{u}\!\cdot\!\vec{v}'}
{1+\gamma_{-}(1+\vec{a}\!\cdot\!\vec{v}')+
g_{-}\vec{u}\!\cdot\!\vec{v}'} ,
\label{eq:r72}
\end{equation}
\begin{equation}
1+\vec{a}\!\cdot\!\vec{v}'
 = \gamma
\frac{(1+\vec{a}\!\cdot\!\vec{u})(1+\vec{a}\!\cdot\!\vec{v})-
(1+a^{2})\vec{u}\!\cdot\!\vec{v} }
{1+\gamma_{+}(1+\vec{a}\!\cdot\!\vec{v})-
g_{+}\vec{u}\!\cdot\!\vec{v}}.
\label{eq:r73}
\end{equation}
It follows from the relations that
\begin{equation}
\gamma(\vec{v}) =  \gamma(\vec{v}')
\left[1+\gamma_{-}(1+\vec{a}\!\cdot\!\vec{v}')+
g_{-}\vec{u}\!\cdot\!\vec{v}'\right] ,
\label{eq:r74}
\end{equation}
\begin{equation}
\gamma(\vec{v}') =  \gamma(\vec{v})
\left[1+\gamma_{+}(1+\vec{a}\!\cdot\!\vec{v})-
g_{+}\vec{u}\!\cdot\!\vec{v}\right] .
\label{eq:r75}
\end{equation}
The formulae generalize Eq. (\ref{eq:r34}).
Region of the accessible values of the velocities is given by
the factor $\gamma$.
The boundary of the region is fixed by the condition
$\gamma(\vec{v}) = \infty$.
For a given value of $\vec{a}$, it is an ellipsoid
\begin{equation}
(v_{\parallel}-a)^2 + (1\!+\!a^2)v^2_{\bot} = 1+a^2
\label{eq:r76}
\end{equation}
in the velocity space.
The focus of the ellipsoid is situated into the point $\vec{v}=0$.
The $v_{\parallel}$ and $v_{\bot}$ denote the velocity components
which are parallel and perpendicular to the vector
$\vec{a}$, respectively.
The ellipsoid is invariant under the relativistic transformations
(\ref{eq:r65}) and (\ref{eq:r70}).
In this case of $\vec{u}=(u,0,0)$ and $\vec{a}=(a,0,0)$, the
composition of the velocities has the simple form
\begin{equation}
v_1' = \frac{v_1-u}{1+2au-uv_1}, \ \ \
v_i' = v_i\frac{\sqrt{1+2au-u^{2}}}{1+2au-uv_1}, \
\ \ \ \ \ i=2,3.
\label{eq:r109}
\end{equation}
The inverse relations can be obtained by
the interchange $\vec{v}\leftrightarrow \vec{v}'$ and
$u\leftrightarrow u'$. Using Eq. (\ref{eq:r40}),
they can be written as follows
\begin{equation}
v_1 = \frac{v_1'+u+2auv_1'}{1+uv_1'}, \ \ \
v_i = v_i'\frac{\sqrt{1+2au-u^{2}}}{1+uv_1'}, \
\ \ \ \ i=2,3.
\label{eq:r110}
\end{equation}

\vskip 0.5cm
{\subsection {Energy and momentum}}

Consider a material particle in space-time.
In relativistic mechanics, the position and momentum of the
particle are given by the four-vectors
$x^{\mu}=\{\vec{x},t\}$ and $p^{\mu}=\{\vec{P},E\}$,
respectively.
Let us define an `elementary' particle as an object which
reveals no internal structure at any resolution considered.
We comprehend the notion of elementarity as a relative
concept which relies on the scales we are dealing with.
For the infinite resolution it should be a perfect
point whose trajectory is a fractal curve.
For an arbitrary small but still finite resolution $\varepsilon^{-1}$
the perfect point is approximated by a particle which we
call `elementary' with respect to this resolution.
It is therefore natural to assume that
the concepts of the momentum, energy,  mass and the velocity of the
`elementary' particle have good physical meaning also
at the scales where space-time is expected to break down its
isotropy.

We denote the values of the momentum and the energy of the
elementary particle by ($\vec{P}$ and  $E$) or
($\vec{P}'$ and $E'$) in the reference systems $S$ or $S'$,
respectively. In consistence with the principle of relativity and
the ideas presented above, we search for relations connecting
these quantities.
In order to do that, let us first define associative variables
$\pi^{\mu}=\{\vec{\pi},\pi_0\}$ with the following property.
The components of the variables determined relative to the systems
$S$ and $S'$ transform in the way
\begin{equation}
\pi' = \Pi(\vec{u})\pi ,
\label{eq:r77}
\end{equation}
where
\begin{equation}
\Pi(\vec{u}) =
\left(
\begin{array}{cc}
\delta_{ij}\!+\!g u_iu_j\!-\!\gamma a_iu_j
& -g_{+}u_i\!+\!\gamma_{+}a_i \\
 -\gamma u_j  & 1\!+\!\gamma_{+} \\
\end{array}
\right) .
\label{eq:r78}
\end{equation}
The inverse transformation $\pi = \Pi^{-1}(\vec{u})\pi'$
is given by the matrix
\begin{equation}
\Pi^{-1} (\vec{u}) =
\left(
\begin{array}{cc}
\delta_{ij}\!+g u_iu_j\!+\!\gamma a_iu_j
& +g_{-}u_i\!+\!\gamma_{-}a_i \\
+\gamma u_j & 1\!+\!\gamma_{-} \\
\end{array}
\right)
\label{eq:r79} .
\end{equation}
There exists mutual correspondence between
the transformation matrices $\Pi$ and $D$ given by the matrix
transposition
\begin{equation}
\Pi (\vec{u}) = D^{\dag} (\vec{u}) .
\label{eq:r80}
\end{equation}
According to the relation, the matrix $\Pi$ can be expressed
in the form
\begin{equation}
\Pi(\vec{u}) = A^{-1}_{\pi}(\vec{a})
\Lambda(\vec{\beta}) A_{\pi}(\vec{a})
\label{eq:r81}
\end{equation}
where
\begin{equation}
A^{-1}_{\pi}(\vec{a}) = A^{\dag}_x(\vec{a}) .
\label{eq:r82}
\end{equation}
The group properties of the transformations (\ref{eq:r77}) are
determined with respect to the composition
\begin{equation}
\Omega_{\pi}(\vec{\phi}) \Pi(\vec{v}) =
\Pi(\vec{v}') \Pi(\vec{u}) ,
\label{eq:r83}
\end{equation}
provided the velocities $\vec{u}$, $\vec{v}'$, and
$\vec{v}$ satisfy the relation (\ref{eq:r65}).
Here
\begin{equation}
\Omega_{\pi} (\vec{\phi}) = A^{-1}_{\pi} R(\vec{\phi}) A_{\pi} .
\label{eq:r84}
\end{equation}
We show that Eqs. (\ref{eq:r64}) and (\ref{eq:r83})
are consistent with relation (\ref{eq:r80}).
Let us transpose Eq. (\ref{eq:r64}).
Exploiting the correspondence (\ref{eq:r80}) and using
Eqs. (\ref{eq:r68}), (\ref{eq:r82}), and (\ref{eq:r84}), we can write
\begin{equation}
\Pi(\vec{v}) \Omega_{\pi}(-\vec{\phi}) =
\Pi(\vec{v}) \Omega^{\dag}_x(\vec{\phi}) =
\Pi(\vec{u}) \Pi(\vec{v}') .
\label{eq:r85}
\end{equation}
We apply the transposition operation on Eq. (\ref{eq:r66})
too.
As the matrices $\Lambda$ are invariant under the operation, we
obtain the composition of the parameters $\vec{\beta}_u$ and
$\vec{\beta}_{v'}$ in the mutual reverse order. From the
symmetry reasons, the composition must be of the same
form as Eq. (\ref{eq:r66}). We have therefore
\begin{equation}
R(-\vec{\phi}) \Lambda (\vec{\beta}_w) =
\Lambda (\vec{\beta}_v) R(-\vec{\phi}) =
\Lambda (\vec{\beta}_u) \Lambda (\vec{\beta}_{v'}) .
\label{eq:r86}
\end{equation}
The vector $\vec{\beta}_w$ corresponds to the velocity
$\vec{w}$ according to Eq. (\ref{eq:r62}).
The velocity is given by the formula (\ref{eq:r65}) in which the
velocities $\vec{u}$ and $\vec{v}'$ are mutually interchanged.
Multiplying Eq. (\ref{eq:r86}) by the $A^{-1}_{\pi}$ from the
left and by the $A_{\pi}$ from the right, we get
\begin{equation}
\Omega_{\pi}(-\vec{\phi}) \Pi(\vec{w}) =
\Pi(\vec{v}) \Omega_{\pi}(-\vec{\phi}) .
\label{eq:r87}
\end{equation}
Together with Eq. (\ref{eq:r85}) one has
\begin{equation}
\Omega_{\pi}(-\vec{\phi}) \Pi(\vec{w}) =
\Pi(\vec{u}) \Pi(\vec{v}') .
\label{eq:r88}
\end{equation}
After performing the interchange
$\vec{u}\leftrightarrow\vec{v}'$,
we obtain Eq. (\ref{eq:r83}). It was thus shown that the
composition of two successive transformations
of the variables $\pi$ follows from the composition of
the corresponding transformations of the coordinates and time,
provided their transformation matrices are
connected by the relation (\ref{eq:r80}).

Unlike the transformations of the coordinates and time,
the invariant combination
\begin{equation}
\pi_0^{2}-\vec{\pi}^{2}+2\pi_0\vec{a}\!\cdot\!\vec{\pi}
\label{eq:r89}
\end{equation}
constructed from the variables $\pi$
does not correspond to the metrics (\ref{eq:r43}).
In order to remove this defect we have to determine
the 4-momentum of free particle by means of new variables.
The transformations of such variables should preserve
the same metric invariant as the transformations of their
kinematical counterparts, the coordinates and time.
We show that there exists two sets of the variables
$p^{\mu}_s = \{\vec{P}_s,E\}$, $s=L,R$ defined by the relation
\begin{equation}
\pi = A_s(\vec{a})p_s ,
\label{eq:r90}
\end{equation}
with
\begin{equation}
A_s(\vec{a}) =
\left(
\begin{array}{cc}
\delta_{ij}\!\pm\!\varepsilon_{ijk} a_k & 0 \\
0 & 1 \\
\end{array}
\right) ,
\label{eq:r91}
\end{equation}
which comply the requirement. Here $\varepsilon_{ijk}$ is
the Levi-Civita symbol.
The plus (in the next every upper) sign and the minus (in the
next every lower) sign corresponds to $s=L$ and $s=R$, respectively.
We will regard the variables $p^{\mu}_s$ as the 4-momentum
of an elementary particle in space-time characterized
by the asymmetry $\vec{a}$.
We attribute the first set of the variables ($s=L$) to the
particle which we call left-handed.
The second set ($s=R$) corresponds to the particle
revealing right-handed type of motion.
The relation between the momenta $\vec{P}_s$
and the above considered variable $\vec{\pi}$ reads
\begin{equation}
\vec{\pi} = \vec{P}_s\pm\vec{P}_s\!\times\!\vec{a} , \ \ \ \ \ \ \
\vec{P}_s =
\frac{\vec{\pi}\pm\vec{a}\!\times\!\vec{\pi}+
(\vec{a}\!\cdot\!\vec{\pi})\vec{a}}
{1+a^{2}} .
\label{eq:r92}
\end{equation}
In the context of these definitions, we
introduce the associative variables $\xi^{\mu}_s=\{\vec{\xi}_s,\xi_0\}$,
$s=L,R$, with respect to the coordinates and time by the formula
\begin{equation}
\xi_s = A_s(-\vec{a})x .
\label{eq:r93}
\end{equation}
The transformations of the variables preserve the invariant
\begin{equation}
\xi_0^{2}-\vec{\xi}^{2}_s+2\xi_0\vec{a}\!\cdot\!\vec{\xi}_s .
\label{eq:r94}
\end{equation}
Let us introduce the parameters $\vec{U}_s=d\vec{\xi}_s/d\xi_0$.
They are related to the velocities $\vec{u}$ as follows
\begin{equation}
\vec{U}_s = \vec{u}\mp\vec{u}\!\times\!\vec{a} , \ \ \ \ \ \ \
\vec{u} =
\frac{\vec{U}_s\mp\vec{a}\!\times\!\vec{U}_s+
(\vec{a}\!\cdot\!\vec{U}_s)\vec{a}}
{1+a^{2}} .
\label{eq:r95}
\end{equation}
Exploiting the additional notations
\begin{equation}
G = \frac{g}{1+a^{2}},
\ \ \ \ \ \ \
G_{\pm} = \frac{g_{\pm}}{1+a^{2}} ,
\label{eq:r96}
\end{equation}
the relativistic transformations
of the energy/momentum take the form
\begin{equation}
p'_s = \Delta(\vec{U}_s)p_s ,
\label{eq:r97}
\end{equation}
where
\begin{equation}
\Delta(\vec{U}) =
\left(
\begin{array}{cc}
\delta_{ij}\!+\!G U_iU_j\!-\!G_{-} a_iU_j
& G U^{2} a_i\!-\!G_{+}U_i \\
-\gamma U_j & 1\!+\!\gamma_{+} \\
\end{array}
\right) ,
\label{eq:r98}
\end{equation}
$U^{2}=\vec{U}\!\cdot\!\vec{U}$.
We will skip the index $s$ in the next, where it is
insubstantial.
Equation (\ref{eq:r50}) implies
\begin{equation}
\vec{U}' = -\frac{\vec{U}}{1+2\vec{a}\!\cdot\!\vec{U}},
\label{eq:r99}
\end{equation}
which together with the symmetry properties (\ref{eq:r51}) and
(\ref{eq:r52}) determine the inverse matrix
\begin{equation}
\Delta^{-1}(\vec{U}) =
\Delta(\vec{U}') =
\left(
\begin{array}{cc}
\delta_{ij}\!+\!G U_iU_j\!+\!G_{+} a_iU_j
& G U^{2} a_i\!+\!G_{-}U_i \\
+\gamma U_j & 1\!+\!\gamma_{-} \\
\end{array}
\right) .
\label{eq:r100}
\end{equation}
The transformation matrixes  can be written in the way
\begin{equation}
\Delta(\vec{U}_s) =
A_{ps}^{-1}(\vec{a}) \Lambda(\vec{\beta}) A_{ps}(\vec{a}) ,
\label{eq:r101}
\end{equation}
where
\begin{equation}
A_{ps}(\vec{a})=A_{\pi}(\vec{a})A_s(\vec{a}) =
\frac{1}{\sqrt{1+a^2}}
\left(
\begin{array}{cc}
\delta_{ij}\!\pm\!\varepsilon_{ijk} a_k & -a_i \\
0 & \sqrt{1+a^2} \\
\end{array}
\right)
\label{eq:r102}
\end{equation}
and $\Lambda$ is given by Eq. (\ref{eq:r60}).

The transformations (\ref{eq:r97}) possess group properties.
Let us consider two successive transformations expressed by
the matrices
$\Delta(\vec{U})$ and $\Delta(\vec{V}')$.
The resultant transformation is given by
\begin{equation}
\Omega_p (\vec{\phi})\Delta(\vec{V}) =
\Delta (\vec{V}')\Delta (\vec{U}) ,
\label{eq:r103}
\end{equation}
provided
\begin{equation}
\vec{V} = \frac{\vec{V}' +
\vec{U}\left[\gamma+G_{+}\vec{a}\!\cdot\!\vec{V}'+
G\vec{U}\!\cdot\!\vec{V}'\right]  }
{1+\gamma_{-}+GU^{2}\vec{a}\!\cdot\!\vec{V}'
+G_{-}\vec{U}\!\cdot\!\vec{V}' } .
\label{eq:r104}
\end{equation}
Formula (\ref{eq:r103}) is consequence of Eqs.
(\ref{eq:r66}) and (\ref{eq:r101}).
The matrix $\Omega_p$ has the structure
\begin{equation}
\Omega_p(\vec{\phi}) \equiv
\Omega_{ps}(\vec{\phi}) =
A^{-1}_{ps} R(\vec{\phi}) A_{ps} =
A_s^{-1} \Omega_{\pi}(\vec{\phi}) A_s .
\label{eq:r105}
\end{equation}
The inverse relation to Eq. (\ref{eq:r104}) reads
\begin{equation}
\vec{V}' = \frac{\vec{V} -
\vec{U}\left[\gamma+G_{-}\vec{a}\!\cdot\!\vec{V}-
G\vec{U}\!\cdot\!\vec{V} \right]  }
{1+\gamma_{+}+GU^{2}\vec{a}\!\cdot\!\vec{V}-
G_{+}\vec{U}\!\cdot\!\vec{V}} .
\label{eq:r106}
\end{equation}
The composition rules (\ref{eq:r104}) and (\ref{eq:r106})
are obtained by substituting Eq. (\ref{eq:r95}) into the formulae
(\ref{eq:r65}) and (\ref{eq:r70}), respectively.

The existence of the space-time asymmetry assumed at small scales
leads us to the conclusion that the energy-momentum vectors
and the space-time positions vectors shall not be treated on the
same footing. For a non-zero value of the asymmetry there exist
two sets of the mechanical variables $p^{\mu}_s$, $s=L,R$ and a single set
of the kinematical variables $x^{\mu}$. The variables are defined in
space-time characterized by the metrics (\ref{eq:r43}).
Single sets of the mechanical variables correspond to the
right-handed and left-handed types of motion, respectively.
Both of them have either positive or negative energy.
The sign of the energy is
conserved in whatever reference frame.
The $4$-vectors $x^{\mu}$ and $p^{\mu}_s$ posses different
transformation properties.
While the first obey the transformation formula (\ref{eq:r55}),
the later are transformed according to Eq. (\ref{eq:r97}).
The variables $x^{\mu}$ and $p^{\mu}_s$  are connected
by the matrices $A_s(-\vec{a})$ and $A_s(\vec{a})$ with the associated
quantities $\xi^{\mu}_s$ and $\pi^{\mu}$, respectively.
The relativistic transformations of the $\xi^{\mu}_s$ and $\pi^{\mu}$
preserve the combinations (\ref{eq:r94}) and (\ref{eq:r89}).
In the special case, when the velocity $\vec{u}$ is parallel to
the vector $\vec{a}$, the difference vanishes,
$\vec{u}=\vec{U}_s$ and the transformation matrices have the form
\begin{equation}
D(\vec{u}) =
\Delta(\vec{U}) =
\left(
\begin{array}{cc}
\delta_{ij}\!+\!\gamma_{-} u_iu_j/u^{2} & -\gamma u_i \\
-\gamma u_j & 1\!+\!\gamma_{+} \\
\end{array}
\right) .
\label{eq:r107}
\end{equation}
The inverse matrix reads
\begin{equation}
D^{-1}(\vec{u}) =
\Delta^{-1}(\vec{U}) =
\left(
\begin{array}{cc}
\delta_{ij}\!+\!\gamma_{+} u_iu_j/u^{2} & +\gamma u_i \\
+\gamma u_j & 1\!+\!\gamma_{-} \\
\end{array}
\right) .
\label{eq:r108}
\end{equation}
The transformations (\ref{eq:r36}) - (\ref{eq:r39})
are recovered by putting $\vec{u}=(u,0,0)$ and $\vec{a}=(a,0,0)$.

Let us now consider the invariant
\begin{equation}
p^2 =
\hat{a}_{\mu\nu}p^{\mu}p^{\nu} = E^2-\vec{P}^{2}+
2E\vec{a}\!\cdot\!\vec{P}-
(\vec{a}\!\times\vec{P})^2
\equiv m_0^2
\label{eq:r111}
\end{equation}
of the transformations (\ref{eq:r97}).
This property follows from the relation
\begin{equation}
\Delta^{\dag}(\vec{U}_s) \hat{a} \Delta(\vec{U}_s) = \hat{a} =
A^{\dag}_{ps}\eta A_{ps}(1\!+\!a^2) .
\label{eq:r112}
\end{equation}
The invariant (\ref{eq:r111}) is proportional to the constant $m_0$, which
is the rest mass assigned to a particle in the non-fractal
and non-relativistic mechanics.
The invariant implies the dependence of the energy of the
particle on its momentum in the following way
\begin{equation}
E = \sqrt{(1+a^2)\vec{P}^2+m^2_0}-\vec{a}\!\cdot\!\vec{P} .
\label{eq:r113}
\end{equation}
We will not consider here the solution with minus sign before
the square root corresponding to anti-particles.
The energy  (\ref{eq:r113}) is positive for arbitrary
values of $\vec{a}$ and $\vec{P}$. It has a single minimum
for the momentum and energy
\begin{equation}
\vec{P}_0 = M_0\vec{a} , \ \ \ \ \ \ \
E(\vec{P}_0) = M_0 .
\label{eq:r114}
\end{equation}
The mass $M_0$ (the minimal energy) depends on the asymmetry
parameter $\vec{a}$ by the relation
\begin{equation}
M_0 = \frac{m_0}{\sqrt{1+a^2}} .
\label{eq:r115}
\end{equation}
Beyond the minimum, as the momentum increases, the energy tends to
infinity. It consists of two terms. The first term
is the free energy
\begin{equation}
{\cal E}
= \sqrt{\vec{P}^2+M^2_0}
\label{eq:r116}
\end{equation}
of an object with the mass $M_0$ scaled by the factor
$(1+a^2)^{1/2}$. The second term,
$V=-\vec{a}\!\cdot\!\vec{P}$, plays the role of a potential
induced by the asymmetry of space-time.
How can we interpret such a result?
The answer was suggested by Nottale. The above energy $E$ includes,
a priori, the potential energy contained in the scale
structure involved. The structure is already present even in the
rest frame of the particle; the rest mass $m_0$ being
itself a geometrical fractal structure of the particle trajectory.
The particle may be identified with its own trajectory, which
is the fractal-like trajectory of a point-like `elementary'
object moving chaotically with the momentum $\vec{P}$
and having the mass (minimum
energy) $M_0(\vec{a}) = E(\vec{P}_0)$.
The chaotic nature of the motion is given by the scale
dependent fluctuations of the parameter $\vec{a}$.
The centre-of-mass frame for the system  consisting of the
chaotically
moving `elementary' object with the scale dependent mass
$M_0(\vec{a})$ is defined by the condition $\vec{P}=0$.
The system represents the counterpart of the
`elementary' object which is the `dressed' particle with the mass
$m_0=E(\vec{P}=0)$.
Really, the relations (\ref{eq:r114}) and (\ref{eq:r115}) can be
inverted and the particle mass
\begin{equation}
m_0 = E_{min}\sqrt{1+a^{2}}
\label{eq:r117}
\end{equation}
is expressed  in terms of the minimal energy
$E_{min}=E(\vec{P}_0)$ and the parameter $\vec{a}$.
We conjecture that similar considerations concern also other
intrinsic characteristics of the particles, such as spin and charge.
One can consider the physical quantities as related to the geometrical
structures of particle trajectories in the fractal space-time.
We anticipate that spin of a particle may be connected to
special erratic character of the left-handed or right-handed
fractal-like trajectory at small scales.
In the domain, where the fractal attributes of the motion expire,
the  value of $\vec{a}$ diminishes and the fractal dynamics
will convert into the relativistic dynamics in smooth space.

We make some comments on the energy momentum conservation.
Let us consider a closed system with the mass $m_0$
which splits into two parts.
The decay is governed by the energy momentum conservation,
$m_0 = \sqrt{\vec{q}^{\ast 2}_1+m^{2}_1}+
\sqrt{\vec{q}^{\ast 2}_2+m^{2}_2}$ and
$\vec{q}^{\ast}_1 = -\vec{q}^{\ast}_2$,
as described in the system rest frame.
The similar is valid in space-time with broken isotropy.
Denoting the energy momentum four-vectors of the decay products
by $p_1$ and $p_2$, one can write
\begin{eqnarray}
m^2_0 &=& (p_1+p_2)^2 =
m^2_1+m^2_2+2p_1p_2 \nonumber \\
&=& \left[E^2_1-\vec{P}^2_1+2E_1\vec{a}\!\cdot\!\vec{P}_1
     -(\vec{a}\!\times\!\vec{P}_1)^2\right] +
    \left[E^2_2-\vec{P}^2_2+2E_2\vec{a}\!\cdot\!\vec{P}_2
     -(\vec{a}\!\times\!\vec{P}_2)^2\right]
\nonumber \\
& &+2\left[E_1E_2-\vec{P}_1\!\cdot\!\vec{P}_2+
E_1\vec{a}\!\cdot\!\vec{P}_2+E_2\vec{a}\!\cdot\!\vec{P}_1
-(\vec{a}\!\times\!\vec{P}_1)(\vec{a}\!\times\!\vec{P}_2)
\right]
\nonumber \\
&=& (E_1+E_2)^2-(\vec{P}_1+\vec{P}_2)^2+
2(E_1+E_2)\vec{a}\!\cdot\!(\vec{P}_1+\vec{P}_2)
-\left[\vec{a}\!\times\!(\vec{P}_1\!+\!\vec{P}_2)\right]^2.
\label{eq:r118}
\end{eqnarray}
We see that if the four vectors $p_1$ and $p_2$ are characterized
by the invariant (\ref{eq:r42}), their sum $p_1+p_2$ possesses
this property too.
This implies the conservation of the total energy and momentum,
$E=E_1+E_2$ and $\vec{P}=\vec{P}_1\!+\!\vec{P}_2$,
which results in the conservation of the free energy
\begin{equation}
\sqrt{\vec{P}^2+M^2_{0}} =
\sqrt{\vec{P}^2_1+M^2_1} +
\sqrt{\vec{P}^2_2+M^2_2}
\label{eq:r119}
\end{equation}
as well. In the center-of-mass system,
the decay is characterized by the products with equal
anti-parallel momenta ($\vec{P}^{\ast}_1=-\vec{P}^{\ast}_2$)
for arbitrary value of $\vec{a}$.
For a given $m_0$, the momenta  are smaller with respect to their
values $\vec{q}^{\ast}_i$ when $\vec{a}=0$.

\vskip 0.5cm
{\subsection {Relations of the kinematical and mechanical
variables}}

Fundamental concepts of the special theory of relativity lead us
to the relation between the energy/momentum of a material
particle and its velocity. The velocity is limited within the sphere
of the radius $c$ in every system of reference and is oriented in
the direction of the particle momentum. This concerns the
description of physical events in the homogeneous and isotropic
space-time where the relativistic principle requires all systems of
inertia to be treated equivalently. We show how the
relations change when we abandon the space-time isotropy
which we expect to break down at small scales.

First we have to determine how depend the variables
$\pi^{\mu}$ on the velocity $\vec{v}$.
In other words, we are searching for the
functions $\vec{f}_1$ and $f_2$,
\begin{equation}
\vec{\pi} = \vec{f}_1(\vec{v},\vec{a}), \ \ \ \ \
\pi_0  = f_2(\vec{v},\vec{a}),
\label{eq:r120}
\end{equation}
which are form invariant with respect to the relativistic
transformations of the variables $\pi^{\mu}$ and the velocity $\vec{v}$.
One can convince itself that the expressions
\begin{equation}
\vec{\pi} =
\left[(1+a^{2})\vec{v} +
(1+\vec{a}\!\cdot\!\vec{v})\vec{a}\right]
\gamma(\vec{v})\frac{m_0}{\sqrt{1+a^{2}}} ,
\label{eq:r121}
\end{equation}
\begin{equation}
\pi_0  = (1\!+\!\vec{a}\!\cdot\!\vec{v})
\gamma(\vec{v})\frac{m_0}{\sqrt{1+a^{2}}}
\label{eq:r122}
\end{equation}
fulfill the requirements. Really, substituting the expressions
into the transformation formula (\ref{eq:r77}), one arrives at
the system
\begin{eqnarray}
\lefteqn{
\left(
\begin{array}{c}
(1\!+\!a^2)v_i' + (1\!+\!\vec{a}\!\cdot\!\vec{v}')a_i   \\
1+\vec{a}\!\cdot\!\vec{v}'  \\
\end{array}
\right)\gamma(\vec{v}') =
}
\\ & &
\hspace*{-5mm} =\left(
\begin{array}{cc}
\delta_{ij}\!+\!g u_iu_j\!-\!\gamma a_iu_j
& -g_{+}u_i\!+\!\gamma_{+}a_i \\
-\gamma u_j & 1\!+\!\gamma_{+} \\
\end{array}
\right)
\left(
\begin{array}{c}
(1\!+\!a^2)v_i + (1\!+\!\vec{a}\!\cdot\!\vec{v})a_i  \\
1+\vec{a}\!\cdot\!\vec{v}  \\
\end{array}
\right)\gamma(\vec{v})
\nonumber
\end{eqnarray}
consisting of four equations. The last one is an identity.
This can be shown by using Eqs. (\ref{eq:r73}) and
(\ref{eq:r75}). In the same manner one can convince itself, that
the first three equations of the system are consistent with Eq.
(\ref{eq:r70}).
Now we substitute Eq. (\ref{eq:r90}) into the formulae
(\ref{eq:r121}) and (\ref{eq:r122}) and get
\begin{equation}
\vec{P}_s = M\left[\vec{v}+
(1\!+\!2\vec{a}\!\cdot\!\vec{v})\vec{a}\right]
\mp M(\vec{v}\!\times\!\vec{a}) ,
\label{eq:r124}
\end{equation}
\begin{equation}
E = (1\!+\!\vec{a}\!\cdot\!\vec{v})M .
\label{eq:r125}
\end{equation}
These are the expressions for the momentum and the energy
of a left-handed ($s\!=\!L$ with upper sign) and right-handed
($s\!=\!R$ with lower sign) `elementary' particle moving
with the velocity $\vec{v}$ in space-time characterized
by the vector anisotropy $\vec{a}$.
As can be seen by direct calculation, the formulae
are consistent with the invariant (\ref{eq:r42}) and
(\ref{eq:r111}).
The coefficient of the proportionality
between the momentum $\vec{P}_s$ and the velocity $\vec{v}$
is denoted by the symbol $M$ and
represents the inertial mass of the particle.
The inertial mass depends on the velocity in the way
\begin{equation}
M(\vec{v}) = M_0\gamma(\vec{v}) .
\label{eq:r126}
\end{equation}
The $M_0$ is the rest mass of the `elementary' particle
given by Eq. (\ref{eq:r115}).
The rest mass corresponds to the minimal energy (\ref{eq:r114}).

Let us now derive the inverse expression with respect to
Eq. (\ref{eq:r124}).
Besides the invariant (\ref{eq:r42}) and (\ref{eq:r111}),
one can construct the invariant relation
\begin{eqnarray}
(A^{\dag}_x\eta A_{ps})_{\mu\nu}x^{\mu}p^{\nu} =
tE-\vec{x}\!\cdot\!\vec{P}_s
+2\vec{a}\!\cdot\!\vec{x}E
\mp\vec{a}\!\cdot\!(\vec{x}\!\times\!\vec{P}_s)
= \tau M_0 .
\label{eq:r127}
\end{eqnarray}
It represents the equation of the elementary particle trajectory expressed
in terms of its momentum, energy, and its mass $M_0$.
The solution of the equation is $\vec{x} = \vec{v}t$,
where
\begin{equation}
\vec{v} = \frac{\vec{P}_s\pm\vec{P}_s\!\times\!\vec{a}-\vec{a}E}
{(1+2a^{2})E-\vec{a}\!\cdot\!\vec{P}_s}.
\label{eq:r128}
\end{equation}
Really, if we substitute the solution into Eq. (\ref{eq:r127})
and exploit the formulae (\ref{eq:r113}), (\ref{eq:r124}), and
(\ref{eq:r125}), we get
\begin{equation}
t=\tau \gamma .
\label{eq:r129}
\end{equation}
The proportionality relates  the time $t$ recorded in a
system $S$ to the particle's proper time $\tau$.
Equation (\ref{eq:r128}) can be rewritten in a more convenient
form
\begin{equation}
M\vec{v} =
\frac{\vec{P}_s\pm\vec{P}_s\!\times\!\vec{a}-\vec{a}E}{1+a^{2}}.
\label{eq:r130}
\end{equation}
We see that for the zero value of the momentum there exists the
non-zero value of the velocity
\begin{equation}
\vec{v}_0 =
-\frac{\vec{a}}{1+2a^2}.
\label{eq:r131}
\end{equation}
The velocity corresponds to the energy $E=m_0$.
According to Eq. (\ref{eq:r111}), the constant $m_0$ is
an invariant of the relativistic transformations and
represents the rest mass assigned to a particle in
the non-fractal (and non-relativistic) limit.
Let us stress that, for the energy $E=m_0$, there exist
infinite number of the mutually correlated values of
$\vec{P}_s$ and $\vec{v}$. They are the momenta and
the velocities attributed to the object
moving along a fractal-like trajectory and representing the
internal structure of the particle itself. In this way, the particle
is reduced to and identified with its own trajectory \cite{Nottale}.
For a given resolution, we identify the object  with
the `elementary' particle possessing the rest mass $M_0$.
The mass depends on the resolution and is
defined in terms of the fractal-like curves characterized
in a purely geometrical way. According to Eq. (\ref{eq:r115}),
its typical value is
\begin{equation}
M_0(<\!a^{2}\!>)= \frac{m_0}{\sqrt{1+<\!a^{2}\!>}} .
\label{eq:r132}
\end{equation}
The symbol $<\!a^{2}\!>$ stands for the average square
of the space-time asymmetries revealed at the considered
level of the resolution.
Other important values of the momentum and velocity giving the
energy $E=m_0$ are $\vec{P}_a=2m_0\vec{a}$
and $\vec{v}_a= \vec{a}$, respectively.
The velocity minimizes the factor $\gamma$.
As shown in the Appendix B,
it corresponds to the minimal length contractions and to the
minimal dilatations of time.

In general, for arbitrary energy of a particle,
we have shown the following result. It consists of the claim
that in space-time with broken isotropy
the momentum of the particle is not parallel to its
velocity. Approximating the fractal space-time by a family of the
spaces $R_{\varepsilon}$ with differentiable geometry,
the velocity fluctuates with respect to
the momentum in dependence on the stochastic nature
of the anisotropy parameter $\vec{a}$.
According to the fluctuations, the `point-like' particle moves
around its momentum passing an unpredictable and chaotic trajectory
characteristic for fractals.
In dependence on the fluctuating anisotropy $\vec{a}$,
the velocity of the particle can be arbitrary large.
This is connected with a possibility of
propagation of physical signals with velocities exceeding
the speed of light in the isotropic space-time.
The property is, however, compensated by the extreme irregular and
random shape of the trajectories along which the signal is mediated.
We stress here that the statements are relative and depend on the
scale of the observer as well. When `measuring' the fractal
properties of the particle motion, the observer expresses them
in terms of its own fractal characteristics being a fractal itself.
This is typical for the parameter $\vec{a}$ which is
a function of the scale structures of both the observed particle
and the observer (see section V.).
According to our opinion, the parameter could have relevance to
more deeper context of the metric potentials which have relation to
the intimate structure of space-time.
It may be connected with a `field of the space-time asymmetry'
reflecting the structure at small scales.
Existence of the `field' would result into a disparity
between the energy-momentum and the coordinates and time.
Here the disparity is demonstrated by the following commutation
relation
\begin{equation}
A_{ps}^{\dag}\eta A_x - A^{\dag}_x\eta A_{ps} =
\left(
\begin{array}{cc}
\pm\epsilon_{ijk}2a_k &  -2a_i  \\
2a_j &  0 \\
\end{array}
\right) .
\label{eq:r133}
\end{equation}
The commutator is non-zero provided the non-zero value of the
field.
In the present paper we approximate the field of the space-time
asymmetry in terms of the anisotropy vector $\vec{a}$ and consider
it as a random and chaotic quantity.
As shown in section V., the anisotropy has relevance to the
anomalous excess of the topological dimensions.
The investigations in this direction require, however, more detailed
and fundamental study.

The ideas tackled in this section concern geodesic
reference systems in the immediate surroundings of a given
point $P$ in the 4-space.
The surroundings depend on the resolution we are dealing with.
One can introduce such systems in the proximity
of every point of the geodesical lines.
According to the ideas about fractal properties of
space-time at small scales, we characterize the geodesic
systems of inertia by the metric tensors $\hat{a}$.
The metrics reflects significant property of
the fractal structure of space-time which is breaking
its isotropy. The structure is revealed in dependence on
the level of the resolution. For a given resolution, it is
possible to transform away the anisotropy of the space-time locally,
exploiting new pseudo-Cartesian coordinates
$r^{\mu}=\{\vec{r},r_0\}$ and $k^{\mu}=\{\vec{K},K_0\}$.
We can introduce the variables in the way
\begin{equation}
r = A_xx, \ \ \ \ \ \
k = A_{ps}p_s .
\label{eq:r134}
\end{equation}
The explicit form of the equations reads
\begin{equation}
\vec{r} = \sqrt{1+a^{2}}\vec{x} ,
\ \ \ \ \ \
r_0 = t+\vec{a}\!\cdot\!\vec{x} ,
\label{eq:r136}
\end{equation}
\begin{equation}
\vec{K} = \frac{1}{\sqrt{1\!+\!a^2}}
\left[\vec{P}_s\pm(\vec{P}_s\!\times\!\vec{a})-\vec{a}E
\right] ,
\ \ \ \ \ \ \ \
K_0 = E .
\label{eq:r135}
\end{equation}
Unlike the $x$ and $p$ the pseudo-Cartesian variables $r$ and $k$
are functions of the anisotropy $\vec{a}$.
Using the variables, one can write the corresponding
relativistic invariant in the form
\begin{equation}
r^2_0(\vec{a})-
\vec{r}^2(a) = \tau^{2},  \ \ \ \ \ \ \ \ \
K^2_0-
\vec{K}^2(\vec{a}) = M^2_0(a) .
\label{eq:r137}
\end{equation}
The space-time anisotropy is thus removable locally but
cannot be removed completely, i.e. simultaneously for every
point of the 4-space. Hence, we consider the anisotropy at small
scales to be the intrinsic property of space-time itself.
Its adequate description assumes approaches within
a fractal geometry.

\vskip 0.5cm
{\section {Interactions of asymmetric fractal systems}}

The ability of fractals to structure space-time was discussed in
Ref. \cite{Nottale}. Such approach gives us possibility to
attribute geometrical notions to the structural parameters
characterizing fractal trajectories of free particles.
We consider one of the parameters to be the scale dependent
coefficient $\vec{a}$ reflecting breaking of the space-time
isotropy.
The quantity is assumed to have stochastic and irregular nature
representing the fractal properties of the structures at small
distances.
The natural question arises whether one can organize a region in
which the structures could be somehow oriented.
We answer the question positively and argue that such region
could be created in the interactions of hadrons and nuclei.
This concerns high energies where the objects reveal fractal
composition in terms of the parton content involved.
The fractality results from non existence of lower cutoff
at which the structures would stop.
We conjecture that the interactions of the fractal objects
affect the character of space-time at small scales.
One can imagine that the chaotic character of the space-time
anisotropy can be oriented and space-time `polarized'
by the interactions of fractals possessing mutually different
anomalous dimensions. In other words, we conjecture that the
interactions of the asymmetric fractal systems result in polarization
of the (QCD) vacuum.
The vacuum fluctuations become oriented forming a
region of the space-time asymmetry.
We denote the asymmetry corresponding to the region
by the vector $\vec{\bar{a}}$. Without the organization, the
parameter represent scale dependent random quantity $\vec{a}$.
As we will show, the $\vec{\bar{a}}$ can be connected with the
anomalous (fractal) dimensions of the interacting fractals.

Let us consider the collision of the asymmetric fractal objects.
The need to satisfy the principles of the scale-motion relativity
implies replacement of the scale
independent physical laws by the scale dependent equations. This
concerns the energy and momentum which in the presence of a
space-time anisotropy are converted to the variables satisfying
the formula (\ref{eq:r113}).
We apply the formula to the relations connecting the
variables of the recoil particle with the corresponding momentum
fractions in the constituent interaction.
We infer on fractal character regarding the motion of the particle
form the requirements which lead to the relations.
They result from the phenomenological analysis of the
$z$ scaling variable and concern the minimal
resolution $\varepsilon^{-1}$ with which one can single out the
constituent interaction underlying the production of the
inclusive particle $m_1$.
The assumption is reflected by the form of the momentum fractions
$\chi_1$ and $\chi_2$ which follows from the condition for
the maximum of the coefficient (\ref{eq:r4}).
According to the requirement, the recoil particle has the energy
$E'$ expressed in the way
\begin{equation}
\frac{2E'}{\sqrt{s}} = \chi_1+\chi_2
= \sqrt{\omega^{2}_1+\mu^{2}_1} +
\sqrt{\omega^{2}_2+\mu^{2}_2} -\left(\omega_1-\omega_2\right) .
\label{eq:f1}
\end{equation}
For the sake of simplicity, all masses $m_i$ and $M_i$ are neglected.
We identify the energy $E'$ with the
expression (\ref{eq:r113}). This gives
\begin{equation}
\sqrt{(1+\bar{a}^{2})(\chi^{2}_z + \chi^{2}_{\bot})}
-\bar{a}\chi_z =
\sqrt{\omega^{2}_1+\mu^{2}_1} +
\sqrt{\omega^{2}_2+\mu^{2}_2} -\left(\omega_1-\omega_2\right) .
\label{eq:f2}
\end{equation}
Here we have used the notations
\begin{equation}
\chi_z = \frac{2P_z}{\sqrt{s}} = \frac{P_z}{E'_{max}} , \ \ \
\chi_{\bot} =
\frac{2P_{\bot}}{\sqrt{s}} = \frac{P_{\bot}}{E'_{max}} .
\label{eq:f3}
\end{equation}
The symbol $\vec{P}$ represents the momentum  of the recoil particle
defined relative to the space-time domain of the elementary
interaction.
Its longitudinal  and transversal components with respect to the
collision axis are denoted by $P_{z}$ and $P_{\bot}$, respectively.
In the collision of asymmetric fractal systems, we characterize
the domain by the space-time anisotropy
$\vec{\bar{a}}=(0,0,\bar{a})$.
As follows from the conservation of the free energy
(\ref{eq:r119}), Eq. (\ref{eq:f2}) splits into two parts
\begin{equation}
\sqrt{(1+\bar{a}^{2})(\chi^{2}_z + \chi^{2}_{\bot})} =
\sqrt{\omega^{2}_1+\mu^{2}_1} +
\sqrt{\omega^{2}_2+\mu^{2}_2} ,
\label{eq:f4}
\end{equation}
\begin{equation}
\bar{a}\chi_z = \omega_1-\omega_2 .
\label{eq:f5}
\end{equation}
The obtained system for the unknown variables $\chi_{z}$ and
$\chi_{\bot}$ depends on the parameter $\bar{a}$.
The variation range of the variables is given by the
condition $\chi_1+\chi_2\le 1$. According to Eqs. (\ref{eq:f1})
and (\ref{eq:f2}), it can be rewritten as follows
\begin{equation}
(\chi_z-\bar{a})^2 + (1\!+\!\bar{a}^2)\chi^2_{\bot} \le
1\!+\!\bar{a}^2 .
\label{eq:f6}
\end{equation}
The $\chi_z$ and $\chi_{\bot}$ are bounded inside the
ellipsoid given by the asymmetry $\bar{a}$.
If we approach the phase-space limit of the reaction
(\ref{eq:r1}), the variables tend to their boundary values
\begin{equation}
\chi_z \rightarrow \tilde\chi_z =
\frac{P^{max}_z}{E'_{max}} , \ \ \
\chi_{\bot} \rightarrow \tilde\chi_{\bot} =
\frac{P^{max}_{\bot}}{E'_{max}} ,
\label{eq:f6a}
\end{equation}
and satisfy the equation of the ellipsoid.
Similar applies for any other particle produced in the elementary
interaction.
The particle's momentum $\vec{P}$ and energy $E'$ are
connected by the dispersion relation (\ref{eq:r113}). In the zero mass
approximation, the relation can be expressed in the way
\begin{equation}
\left(\frac{P_z}{E'}-\bar{a}\right)^2
+ (1\!+\!a^2)\left(\frac{P_{\bot}}{E'}\right)^2
= 1+\bar{a}^2 .
\label{eq:f7}
\end{equation}
As follows from Eqs. (\ref{eq:f1}) and (\ref{eq:f2}), it is
identical to the equation
\begin{equation}
\left(\frac{\chi_z}{\chi_1+\chi_2}-\bar{a}\right)^2
+ (1\!+\!a^2)\left(\frac{\chi_{\bot}}{\chi_1+\chi_2}\right)^2
= 1+\bar{a}^2 ,
\label{eq:f8}
\end{equation}
where
\begin{equation}
\frac{P_z}{E'} =
\frac{\sqrt{s}}{2E'}\chi_z =
\frac{\chi_z}{\chi_1+\chi_2} , \ \ \ \ \ \
\frac{P_{\bot}}{E'} =
\frac{\sqrt{s}}{2E'}\chi_{\bot} =
\frac{\chi_{\bot}}{\chi_1+\chi_2} .
\label{eq:f9}
\end{equation}
The values of $\chi_z/(\chi_1\!+\!\chi_2)$ are limited
within the interval
\begin{equation}
\bar{a}_{-}\le\frac{\chi_z}{\chi_1+\chi_2}\le \bar{a}_{+}
\label{eq:f10}
\end{equation}
with
\begin{equation}
\bar{a}_{\pm} = \bar{a} \pm \sqrt{1+\bar{a}^{2}}.
\label{eq:f11}
\end{equation}
According to the kinematics of the process, the maximal value
of $\chi^{max}_z/(\chi_1\!+\!\chi_2)=\bar{a}_{+}$ should correspond
to $\chi_2=0$. The minimal value of
$\chi^{min}_z/(\chi_1\!+\!\chi_2)=\bar{a}_{-}$ is given
by $\chi_1=0$.
The maximum of $\chi_{\bot}/(\chi_1\!+\!\chi_2)=1$
should be achieved for
$\chi_1=\chi_2$ and thus for $\chi_z/(\chi_1+\chi_2)=\bar{a}$.
The conditions are satisfied by the linear combination
\begin{equation}
\chi_z =
(\chi_1+\chi_2)\bar{a}+(\chi_1-\chi_2)\sqrt{1+\bar{a}^{2}}.
\label{eq:f12}
\end{equation}
Substituting the expression (\ref{eq:f12}) into the relation
(\ref{eq:f5}), one arrives at the equation for the asymmetry $\bar{a}$.
Its solution which complies the physical requirements on the
kinematics of the subprocess reads
\begin{equation}
\bar{a} = \frac{\alpha-1}{2\sqrt{\alpha}}\lambda_c ,
\label{eq:f13}
\end{equation}
where
\begin{equation}
\lambda_c = \sqrt{\frac{\lambda_1\lambda_2}
{(1-\lambda_1)(1-\lambda_2)}} , \ \ \ \ \ \lambda_c \le 1.
\label{eq:f14}
\end{equation}
Using Eqs. (\ref{eq:r9}), (\ref{eq:r10}), (\ref{eq:f5}), (\ref{eq:f8})
and (\ref{eq:f13}),
one can express the variables $\chi_z$ and $\chi_{\bot}$
in a simple form
\begin{equation}
\chi_z = \mu_1-\mu_2 , \ \ \ \ \ \ \ \
\chi_{\bot} = 2\sqrt{\mu_1\mu_2} .
\label{eq:f15}
\end{equation}
When approaching the phase-space limit, $\lambda_c\rightarrow 1$
and the value of $\bar{a}$ becomes maximal.
In the extreme case, the fractions $\lambda_i$ approach
their boundary values
\begin{equation}
\lambda_1\rightarrow \tilde\lambda_1=\cos^2(\theta/2) ,
\ \ \ \ \ \ \ \ \ \
\lambda_2\rightarrow \tilde\lambda_2=\sin^2(\theta/2) .
\label{eq:f16}
\end{equation}
Here $\theta$ is the detection angle of the inclusive particle $m_1$.
As seen from Eqs. (\ref{eq:r8}) - (\ref{eq:r11}), this
corresponds to the transition
\begin{equation}
\chi_z \rightarrow \tilde\chi_z =
\sqrt{\alpha}\sin^2(\theta/2)-
\frac{1}{\sqrt{\alpha}}\cos^2(\theta/2),
\ \ \ \ \ \ \
\chi_{\bot} \rightarrow \tilde\chi_{\bot} = \sin\theta .
\label{eq:f17}
\end{equation}

All the expressions are given in terms of the
coefficient $\alpha$ which is the ratio of the anomalous fractal
dimensions of the colliding objects.
The collisions of the asymmetric fractal systems are
characterized by the different fractal dimensions and thus with
$\alpha \ne 1$.
In the considered scenario, it results in creation of the
domain in which the isotropy of space-time is violated.
The space-time anisotropy in the interaction region is given
by the formula (\ref{eq:f13}).
If $\alpha=1$, there is no polarization of space-time
induced by the interaction.
This corresponds to the collisions of the fractals
possessing equal fractal dimensions.  Similar situation concerns
the interaction of the objects which reveal no fractal-like
substructure.
The asymmetry $\bar{a}$ becomes non-zero for $\alpha\ne 1$.
It changes its sign if $\lambda_1\leftrightarrow\lambda_2$ and
$\alpha\leftrightarrow\alpha^{-1}$, i.g. if the interacting fractals
are mutually interchanged.
The parameter $\bar{a}$ is the product of the induced asymmetry
\begin{equation}
\bar{a}_0 = \frac{\alpha-1}{2\sqrt{\alpha}}
\label{eq:f18}
\end{equation}
and the factor $\lambda_c$. The induced asymmetry of
space-time results from the interaction of the fractals
characterized by mutually different anomalous (fractal) dimensions.
The value of the asymmetry was identified \cite{Z99} with
the space component of the four velocity
\begin{equation}
\frac{{\cal V}}{\sqrt{1-{\cal V}^{2}}} = \bar{a}_0.
\label{eq:f19}
\end{equation}
The velocity ${\cal V}$ has its origin in the asymmetry of the
interaction and vanishes in the collisions of objects which
possess equal fractal structures $(\alpha=1)$.
It can be expressed by the form
\begin{equation}
{\cal V} = \frac{\alpha-1}{\alpha+1}
\label{eq:f20}
\end{equation}
representing the velocity of a `space-time drift'
induced by the interaction of the parton fractals.
The quantity represents no real motion but characterizes
local polarization of the vacuum.
The velocity depends on the state of scale of the reference
systems and satisfies the scale-relativity composition rule
\begin{equation}
{\cal V} =
\frac{{\cal V}_1+{\cal V}_2}{1+{\cal V}_1{\cal V}_2} ,
\label{eq:f21}
\end{equation}
provided
\begin{equation}
\alpha = \alpha_1\alpha_2.
\label{eq:f22}
\end{equation}
If we exploit the experimentally established \cite{Z99} relation
$\delta_A=A\delta$, the last equation can be rewritten
as follows
\begin{equation}
\frac{A_3}{A_1} = \frac{A_3}{A_2}\frac{A_2}{A_1}.
\label{eq:f23}
\end{equation}
This means that the state of scale of the reference system
possesses natural scaling property consisting in the following.
If one examines the nucleus $A_3$ by means of the probe $A_2$,
and then the probe nucleus $A_2$ by another probe $A_1$,
one arrives at the similar structure as if examining the nucleus
$A_3$ with the probe $A_1$.
Physically, the structures are characterized by the asymmetry
in the interactions of the fractals which occurs as a consequence of
the richer parton content of one fractal as compared to the other
one.

The second factor in Eq. (\ref{eq:f13}), $\lambda_c$,
represents projection of the induced asymmetry onto
the asymmetry value as perceptible from the corresponding
resolution $\varepsilon^{-1}$.
Diminishing the resolution the observed asymmetry $\bar{a}$
decreases.
Let us look at possible manifestations of the
asymmetry from the experimental point of view.
The necessary condition for the polarization of
space-time in the interactions of hadrons and nuclei
is the high energy regime where the interacting objects reveal
internal fractal-like substructure. As one can conjecture from
the analysis of the data on inclusive particle production in view
of the $z$ scaling regularity,
the regime is expected to set on in the energy region
$\sqrt{s_{NN}}\ge 20$ GeV where the $z$ scaling becomes valid.
In order to deal with sufficient asymmetry,
we have to consider the processes in which the factor $\lambda_c$
is large enough.
This concerns the interactions with large transverse
momenta of the observed secondaries.
We have estimated the expected asymmetry in the case of the
experimentally measured inclusive reactions \cite{Antre} at 400~GeV
proton incoming energy.
For the $pA$ interactions $\alpha=A$ and
$\lambda_c\simeq E_{\bot}/(\sqrt{s_{NN}}\sqrt{A})$.
The asymmetry was evaluated according to the formula (\ref{eq:f13})
in the most optimistic case of $E_{\bot}=7$~GeV.
We have obtained the values $\bar{a}\sim 0.09\div 0.13$ for
various target nuclei.
The relatively high estimates are rather heuristic
and should not be taken literally.
One a priori does not know whether the full asymptotic regime
with respect to the fractal properties of the interaction is
achieved at the considered center-of-mass energy.
This may occur at higher energies where, for the given transverse
momentum, the projection factor $\lambda_c$ becomes much smaller.
The experimental search
for the effect should thus relay on the detection of the particles
with still higher momenta. It is connected with
difficulties in measurements of small cross sections at which the
particles are observable. This concerns also the statistical
analysis of an experiment from which one could
infer on the existence of the possibility to induce
a polarization of space-time.

\vskip 0.5cm
{\section {Summary}}

The questions addressed in the paper concern general properties
of the particle production at high energies.
The properties are connected with the notions such as locality,
self-similarity and fractality in the collisions of hadrons and
nuclei. They are manifested mostly in the relativistic regime of
local parton interactions which underlie the production of the
observed secondaries.
In this regime, the description of the inclusive cross sections
reveals scaling behavior in dependence on the single variable $z$.

We have discussed some aspects of the relation between
the fractality of the interacting objects and the fractal properties
of space-time.
It is relevant for small scales where the parton composition
of the objects is supposed to reveal a fractal-like substructure.
The assumption has fundamental consequence which is breaking
of the reflection invariance at the infinitesimal level.
Special attention is dedicated to the elaboration of the
formalism concerning the relativity in spaces with broken
isotropy.
Our treatment corresponds to a change in the energy formula
in the relativistic case.
We have obtained explicit relations between the energy/momentum and
the velocity in space-time characterized by the asymmetry
$\vec{a}$. In view of these results, increase of stochasticity
of the parameter with decreasing scales would result
in unpredictable fractal-like motion of
particles with respect to their momenta.
This implies change of the rest mass $M_0$ in dependence on the
value of $\vec{a}$ as well as possibility of motion with
the velocities exceeding the speed of light in isotropic
space-time.
We have determined the coefficient characterizing the anisotropy
of space-time in the interactions of the asymmetric fractal
systems.
It is expressed in terms of the anomalous dimensions of
the fractal objects (hadrons and nuclei) colliding at high
energies.
The relation is illuminated with respect to the choice of the
scaling variable $z$. The variable $z$ represents a fractal measure
proportional to the formation length of a produced particle.
The scaling hypothesis states that the differential cross
section for the production of the particle depends at high energies
on its formation length universally and in an energy independent way.
The evolution of the formation process is expressed by the
scaling function $H(z)$.
The proposed scenario is stressed by the results of our analysis concerning
experimental data at high energies. Namely, based on the confrontation
of the $z$ scaling scheme with the experimental data, we have shown that
the anomalous fractal dimensions for the inclusive production of pions
($\delta\sim 0.8$) and for jets ($\delta\sim 1$) nearly correspond to the
relation $D=1+\delta=2$. The relation characterizes fractal dimension of
Feynman trajectories and is a direct consequence
\cite{Nottale} of the Heisenberg uncertainty relations.

Presented approach to the $z$ scaling shows that the observed
regularity can have relevance to fundamental principles of
physics at small scales.
The general assumptions and ideas discussed here underline
need of searching new approaches to physics at ultra-relativistic
energies.
This concerns better understanding of the micro-physical domain
tested by large accelerators of hadrons and nuclei.

\vskip 1cm

{\Large \bf Acknowledgment}

\vskip 0.5cm

This work has been partially supported
by the Grant of the Czech Academy of
Sciences No. 1048703 (128703).

\vskip 1cm
{\Large \bf Appendix A}
\vskip 0.5cm

We would like to present some properties which follow from the
determination of the variables used in our scheme.
The elementary interaction of constituents is characterized by
the momentum fractions $x_1$ and $x_2$. The relation between the
fractions is given by the minimum recoil mass hypothesis
in the constituent interaction.
The variables are determined in a way to maximize the value of
$\Omega$, which gives the minimal resolution $\varepsilon^{-1}$.
Each interacting constituent consists from a leading part
carrying the momentum fraction $\lambda_i$ and of a parton
`coat' which is a fractal cloud of tiny partons with the momentum
fraction $\chi_i$.
What penetrates the cloud is usually determined
by the virtuality of a probe and is connected with the resolution.
The situation is, however, different as compared to the
deep inelastic processes where the `elementary' interaction is
fixed by the kinematical characteristics of the lepton scattering.
In the collisions of the composite objects such as hadrons and
nuclei, one can, in principle, recognize the interactions
of constituents which underlie the production processes, as well.
The level of recognition is given by the resolution
$\varepsilon^{-1}$, with which one can single out the
corresponding subprocesses. It concerns both hard and soft
collisions characterized by the different momentum transfer. This
in turn determines the virtuality of a probe carrying the momentum
transferred and penetrating into the fractal substructure of the very
constituents.
The squares  of the 4-momenta transferred $-Q_1^2$ and $-Q_2^2$
from the first and the second interacting constituent are as follows
\begin{equation}
Q_1^2=(x_1P_1\!-\!q)^2 , \ \ \ \ \ \ \
Q_2^2=(x_2P_2\!-\!q)^2.
\label{eq:b13}
\end{equation}
In the zero mass approximation, the quantities are correlated
with the square of the subprocess energy $s_x$ via the relation
\begin{equation}
s_x+Q_1^2+Q_2^2=0.
\label{eq:b14}
\end{equation}
The transferred momenta are usually considered as virtualities of
the probes
that penetrate the internal structure of the interacting objects.
If the underlying interaction of constituents does not posses
the contact character, the virtualities are carried by the quanta
of the calibration fields.
Then the fields mediate the interaction between the constituents.
The transferred momenta $-Q_1^2$ and $-Q_2^2$ are
connected with resolution.
They are equal for
\begin{equation}
\frac{\chi_1}{\chi_2}=\frac{\lambda_1}{\lambda_2}.
\label{eq:b15}
\end{equation}
The condition determines the boundary between the phase space
hemispheres \cite{Z99} belonging to the interacting objects 1 and 2.
We have $-Q_1^2>-Q_2^2$ in the hemisphere corresponding to the object 2.
This region is preferable to study of the processes in which
the constituents from the object 1 penetrate deeper into the structure
of the object 2 and test it in more detail.
For $-Q_1^2<-Q_2^2$ it is vice versa.
With the increasing values of $-Q_i^2$, the interaction of the
constituents take place on still smaller distances. This in turn increases
the spatial resolution necessary for investigation of fractality
at small scales.

Next we will show that our determination of the momentum fractions
accounts for the back-to-back topology in the constituent's center-of-mass
system $S_c$.
First, let us consider the momenta of the inclusive particle and its
recoil in the total center-of-mass system $S$.
For the nucleon-nucleon collisions the parameter $\alpha=1$ and the
`coats' of the interacting constituents carry the momentum fractions
\begin{equation}
\chi_1 \rightarrow \bar{\mu}_1 \equiv \lambda
\sqrt{\frac{1\!-\!\lambda_1}{1\!-\!\lambda_2}}, \ \ \ \ \
\chi_2 \rightarrow \bar{\mu}_2 \equiv \lambda
\sqrt{\frac{1\!-\!\lambda_2}{1\!-\!\lambda_1}} .
\label{eq:b1}
\end{equation}
The constituents are indistinguishable for $x_1=x_2$.
In this region, each of them possesses the cloud of tiny partons
with the same momenta given by $\bar{\mu}_1=\bar{\mu}_2$.
This is not longer valid for $x_1\ne x_2$. Let us assume that $x_1> x_2$.
It follows from Eqs. (\ref{eq:r7}) and (\ref{eq:r9}) that
$\lambda_1>\lambda_2$ and $\bar{\mu}_1<\bar{\mu}_2$.
This implies the situation when the recoil object moves in the
direction not precise opposite to the inclusive particle $m_1$ in the
system $S$.
For the sake of simplicity, we demonstrate this statement in
the approximation when all masses are neglected.
We use the notations
\begin{equation}
\lambda_1 =\frac{P_2q}{P_1P_2} \rightarrow
\frac{E+q_z}{\sqrt{s}}, \ \ \ \ \
\lambda_2 =\frac{P_1q}{P_1P_2} \rightarrow
\frac{E-q_z}{\sqrt{s}},
\label{eq:b2}
\end{equation}
\begin{equation}
\bar{\mu}_1 =\frac{P_2\bar{q}}{P_1P_2} \rightarrow
\frac{\bar{E}+\bar{q}_z}{\sqrt{s}}, \ \ \ \ \
\bar{\mu}_2 =\frac{P_1\bar{q}}{P_1P_2} \rightarrow
\frac{\bar{E}-\bar{q}_z}{\sqrt{s}},
\label{eq:b3}
\end{equation}
\begin{equation}
\chi_1 =\frac{P_2q'}{P_1P_2} \rightarrow
\frac{E'+q_z'}{\sqrt{s}}, \ \ \ \ \
\chi_2 =\frac{P_1q'}{P_1P_2} \rightarrow
\frac{E'-q_z'}{\sqrt{s}},
\label{eq:b4}
\end{equation}
introducing the energy and momentum for the recoil particle by
the symbols $\bar{E}$ and $\bar{q}$
(or $E'$ and $q'$ for $\alpha\ne 1$) in the center-of-mass system $S$.
The angles  contained by the momenta $\vec{q}$, $\vec{\bar{q}}$,
and $\vec{q}'$  with the collision axis oriented in the
direction of motion of the colliding object $1$ are given by the
expressions
\begin{equation}
\tan(\theta/2) = \sqrt{\frac{\lambda_2}{\lambda_1}}, \ \ \ \ \ \
\tan(\bar{\theta}/2) = \sqrt{\frac{\bar{\mu}_2}{\bar{\mu}_1}}, \
\ \ \ \ \
\tan(\theta'/2) = \sqrt{\frac{\chi_2}{\chi_1}},
\label{eq:b5}
\end{equation}
respectively.
The relations $x_1>x_2$ ($x_1<x_2$) imply
$\theta+\bar{\theta}<\pi$ ($\theta+\bar{\theta}>\pi$).
This can be proved as follows. Let e.g. $x_1>x_2$. It is
equivalent to $\lambda_1>\lambda_2$ and
\begin{equation}
1> \sqrt{\frac{\lambda_2}{\lambda_1}}
\sqrt{\frac{1-\lambda_2}{1-\lambda_1}}.
\label{eq:b6}
\end{equation}
Exploiting Eqs. (\ref{eq:b1}) and (\ref{eq:b2}), we can write
$1>\tan(\theta/2)\tan(\bar{\theta}/2)$, and consequently
$\theta+\bar{\theta}<\pi$. The inverse inequalities can be proved
equivalently. We have thus shown that in the $2\rightarrow 2$
processes  there is perfect back-to-back correlation
between the inclusive particle an its recoil in the reference
system $S$ only for $x_1=x_2$.
This is valid also for the reactions where the parameter
$\alpha\ne 1$.
Change of the parameter corresponds to a change of the scale
of the reference system and results in changing of the resolution.
In the center-of-mass system $S$, the constituent subprocess reveals
back-to-back topology in the special case
\begin{equation}
\cos\theta = \frac{1-\alpha}{1+\alpha} .
\label{eq:b8}
\end{equation}
The situation corresponds to the equal momentum fractions
$x_1=x_2$.
On the other hand one gets $\theta+\theta'<\pi$ or $\theta+\theta'>\pi$
for $x_1>x_2$ or $x_1<x_2$, respectively.
We will show how the relations change in the constituent's center-of-mass
system $S_c$. For $x_1=x_2$, it coincides with the
total center-of-mass frame $S$.
For $x_1\ne x_2$, the system $S_c$ is determined by the condition
$x_1P^c_1=x_2P^c_2$ where $P^c_i$ are the momenta of the colliding
nuclei in $S_c$. The momenta $P^c_i$ can be related to the
momenta $P_i$ as follows
\begin{equation}
P^c_1=P_1\sqrt{\frac{x_1}{x_2}} , \ \ \ \ \ \
P^c_2=P_2\sqrt{\frac{x_2}{x_1}} .
\label{eq:b9}
\end{equation}
This allows us to write
\begin{equation}
\lambda_1= \frac{P^c_2q^c}{P^c_1P^c_2}=
\frac{E^c+q^c_z}{\sqrt{s}} \sqrt{\frac{x_1}{x_2}}, \ \ \ \ \ \
\lambda_2= \frac{P^c_1q^c}{P^c_1P^c_2}=
\frac{E^c-q^c_z}{\sqrt{s}} \sqrt{\frac{x_2}{x_1}}.
\label{eq:b10}
\end{equation}
The angles $\theta_c$ and $\theta_c'$ contained by the
momenta $\vec{q}_c$ and $\vec{q}_c'$  with the collision axis
can be expressed in the system $S_c$ in the way
\begin{equation}
\tan(\theta_c/2) = \sqrt{\frac{\lambda_2x_1}{\lambda_1x_2}}, \ \ \ \ \ \
\ \ \ \ \
\tan(\theta_c'/2) = \sqrt{\frac{\chi_2x_1}{\chi_1x_2}}.
\label{eq:b11}
\end{equation}
This implies
\begin{equation}
\tan(\theta_c/2)\tan(\theta_c'/2)=1
\label{eq:b12}
\end{equation}
and consequently $\theta_c+\theta_c'=\pi$. Really, the substitution
of expressions (\ref{eq:b11}) into Eq. (\ref{eq:b12})
gives
\begin{equation}
\sqrt{\lambda_2\chi_2}(\lambda_1+\chi_1) =
\sqrt{\lambda_1\chi_1}(\lambda_2+\chi_2) .
\label{eq:b13a}
\end{equation}
It remains to exploit the relation $\lambda_1\lambda_2=\chi_1\chi_2$
and one gets the identity.
We have thus shown that our determination of the momentum fractions
is consistent with back-to-back topology of the collisions
in the center-of-mass
systems of the interacting constituents.

\vskip 1cm
{\Large \bf Appendix B}
\vskip 0.5cm

In this Appendix we discuss some aspects concerning the
relativistic transformations of the energy and the coordinates
in space-time with broken isotropy.
Reasonable definition of the variables assumes the fulfillment
of certain requirements resulting from the proper composition of the
velocities. It regards the principle of causality
and the constraint on the positivity of the energy.
At the end we add some comments on the character of
the lengths contractions and the dilatations of time.

According to the special theory of relativity, the values
of the particle's velocities are bounded  within the sphere
$u \le 1$ in any inertial frame.
This is given by the factor $\gamma$ which for
$\vec{a}=0$ and for the superluminous velocities becomes imaginary.
The situation changes if we admit the breaking of the space-time
isotropy expressed by the non-zero value of $\vec{a}$.
The velocity sphere deforms to an ellipsoid with the focus in the
beginning of the velocity space. Center of the ellipsoid is shifted
into the point $\vec{u}=\vec{a}$ and its larger axis
becomes $\sqrt{1+a^{2}}$.
We illustrate the basic properties of the velocity composition
in one dimensional case ($a\equiv a_1$, $u\equiv u_1$).
The region of the accessible values of the velocities is
determined by the condition
\begin{equation}
a_{-} \le u \le a_{+}
\label{eq:a1}
\end{equation}
where
\begin{equation}
a_{-} = a-\sqrt{1+a^{2}}, \ \ \ \ \ \ \ \
a_{+} = a+\sqrt{1+a^{2}} .
\label{eq:a2}
\end{equation}
The boundaries $a_{-}$ and $a_{+}$ satisfy the relation
$\gamma(a_{\pm})=\infty$.
We have $-1 \le a_{-} \le 0$ and $1 \le a_{+} \le \infty$
for $a\ge 0$.
The negative values of $a$ imply
$-\infty \le a_{-} < -1$ and $0 \le a_{+} < 1$.
Consider two velocities, $u$ and $v'$.
If the velocities are from the interval
$a_{-} \le u,v' \le a_{+}$, the composed velocity (\ref{eq:r33})
is bounded by the condition $a_{-} \le v \le a_{+}$ as well.
If the limiting velocity $a_{-}$ or $a_{+}$ is composed with a
velocity $u$,
\begin{equation}
a_{-} = \frac{a_{-}+u+2aa_{-}u}{1+a_{-}u} , \ \ \ \ \
a_{+} = \frac{a_{+}+u+2aa_{+}u}{1+a_{+}u} ,
\label{eq:a3}
\end{equation}
one gets again $a_{-}$ or $a_{+}$, respectively.
As follows from the relations
\begin{equation}
a_{+} = -\frac{a_{-}}{1+2aa_{-}}  , \ \ \ \
a_{-} = -\frac{a_{+}}{1+2aa_{+}}  ,
\label{eq:a4}
\end{equation}
the limiting velocities $a_{-}$ and $a_{+}$ are mutually inverse
with respect to Eq. (\ref{eq:r40}).
For $a>0$, the instant velocity of the particle is bounded from above
by the value of $a_{+}$ which is larger than unity.
This gives possibility of the motion with the velocities
exceeding the speed of light $c$ in isotropic space-time.

We will show that propagation of an energetic signal with
such velocities fulfills the principle of causality.
According to the requirement, the consequence - the detection of
a signal can not precede its emission in whatever system of
reference.
Let us assume that the signal was emitted in the
point $(x_1,t_1)$ and detected at $(x_2,t_2)$
with respect to the frame $S$, $dt=t_2-t_1> 0$.
The velocity of the signal propagation is $v=dx/dt$,
$dx=x_2-x_1$.
Let us look at the two events from the system $S^{'}$
moving relatively to the initial one with the speed $u$.
According to the transformation (\ref{eq:r37}) we have
\begin{equation}
dt' = \gamma(u) \left[(1+2au)dt-udx\right] =
\gamma(u)dt\left(1+2au-uv\right) .
\label{eq:a5}
\end{equation}
The factor on the right hand side is
non-negative for any velocities from the interval
$a_{-} \le u,v \le a_{+}$.
Consequently $dt'=t_2'-t_1' \ge 0$.
The same is valid in the general case when the signal propagates
between the points $(\vec{x}_1,t_1)$ and $(\vec{x}_2,t_2)$
with the velocity $\vec{v}=d\vec{x}/dt$.
We assume that the system $S'$ is moving with respect to the
reference frame $S$ with the velocity $\vec{u}$.
As follows from Eqs. (\ref{eq:r55}) and (\ref{eq:r56}),
the time interval $dt'$ of the signal propagation relative to
the system $S'$ is given by
\begin{equation}
dt' = (1+\gamma_{+})dt+\gamma_{+}\vec{a}\!\cdot\!d\vec{x}
-g_{+}\vec{u}\!\cdot\!d\vec{x}
= [1+\gamma_{+}(1+\vec{a}\!\cdot\!\vec{v})
-g_{+}\vec{u}\!\cdot\!\vec{v}]dt .
\label{eq:a6}
\end{equation}
We see from Eq. (\ref{eq:r75}) that
the expression in the brackets is non-negative.
This implies $dt'\ge 0$ in agreement with the causality
principle, which is not violated in space-time
with broken isotropy.

The next step is to prove the positivity of the transmitted
energy.
Combining Eqs. (\ref{eq:r121}) and (\ref{eq:r122}), one has
\begin{equation}
\vec{\pi} = \frac{(1+a^{2})}{1+\vec{a}\!\cdot\!\vec{v}}
\pi_0\vec{v}+\pi_0\vec{a} .
\label{eq:a7}
\end{equation}
We insert this expression into Eq. (\ref{eq:r77})
and after some manipulation write
\begin{equation}
E' =
E\frac{\gamma}{1+\vec{a}\!\cdot\!\vec{v}}
\left[(1+\vec{a}\!\cdot\!\vec{u})
(1+\vec{a}\!\cdot\!\vec{v})-
(1+a^{2})\vec{u}\!\cdot\!\vec{v}\right] .
\label{eq:a8}
\end{equation}
When exploiting Eqs. (\ref{eq:r73}) and (\ref{eq:r75}),
the relation can be rewritten into the form
\begin{equation}
E' = E
\frac{(1+\vec{a}\!\cdot\!\vec{v}')\gamma(\vec{v}')}
{(1+\vec{a}\!\cdot\!\vec{v})\gamma(\vec{v})} .
\label{eq:a9}
\end{equation}
As follows from the inequality
\begin{equation}
0 \le 1+aa_{-} \le 1+\vec{a}\!\cdot\!\vec{v}
\label{eq:a10}
\end{equation}
valid for any velocity $\vec{v}$ bounded by the ellipsoid
(\ref{eq:r76}), the factors on the right side of the
relation (\ref{eq:a9}) are non-negative.
The energy of the signal is thus positive in each system of
reference $S'$ which is moving relatively to the system $S$
with the velocity $\vec{u}$.
The above mentioned properties enable the
propagation of physical signals including
transportation of the energy with the velocities
exceeding the value of $c$ - the speed of light in isotropic
space-time. This can occur at small scales
within the regions with broken space-time isotropy.

Based on the transformations (\ref{eq:r55}), we can
draw conclusions regarding the course of time and
the change of lengths of the elementary sections expressed relative
to the systems $S$ and $S'$.
Let us consider a clock at rest with respect to the system $S'$.
Time recorded by the clock is referred as the proper time.
According to Eq. (\ref{eq:r129}), the increase of the
proper time $d\tau$ and the corresponding increase of time $dt$
in the system $S$ are related as follows
\begin{equation}
dt(\vec{v}) =
\frac{d\tau}{\sqrt{(1+\vec{a}\!\cdot\!\vec{v})^2-
(1\!+\!a^2)v^{2}}} .
\label{eq:a11}
\end{equation}
In view of the asymmetry represented by the factor $\vec{a}$, the
course of the clock time can be even faster than in its rest
frame, when observed by a moving observer. The minimal time
interval between two events
\begin{equation}
dt(\vec{v}_a)
= \frac{d\tau}{\sqrt{1+a^{2}}} \le d\tau , \ \ \ \
\vec{v}_a = \vec{a} ,
\label{eq:a12}
\end{equation}
is recorded from the system $S$ in which the clocks are moving
with the velocity $\vec{v}_a$.
The clocks are slowing down (their time intervals increase)
if their velocity approaches the limit given by
Eq. (\ref{eq:r76}).

The change of the length of a section with the velocity
is little more complicated, thought its transverse dimensions
with respect to the motion are not subjected to any change.
Indeed, if the velocity is
oriented e.g. in the direction of the $x$-axis, the $y$ and $z$
components of the coordinates are invariant with respect to the
transformation (\ref{eq:r55}).
As concerns the longitudinal contractions, we discuss here the
simplified situation in which the section has no transverse
dimension relative to its motion.
It is natural to define the length $dl$ of the section with
respect to $S$ as the difference between the
simultaneous coordinate values of its end-points. If the section
is at rest in the $S'$ system, its rest length is given by $dl_0=
x'_2-x'_1$. According to the specific situation considered,
the both values are connected with the expression
\begin{equation}
dl(\vec{v}) = dl_0 \sqrt{(1+\vec{a}\!\cdot\!\vec{v})^2-
(1\!+\!a^2)v^{2}} .
\label{eq:a13}
\end{equation}
The length $dl(\vec{v})$ observed from the system $S$ is maximal
if the section moves with the velocity
$\vec{v}_a=\vec{a}$. Its value depends
on the parameter $\vec{a}$ in the way
$dl(\vec{v}_a) = dl_0\sqrt{1+a^{2}} \ge dl_0$ .
%\begin{equation}
%dl(\vec{v}_a) = dl_0\sqrt{1+a^{2}} \ge dl_0 .
%\label{eq:a14}
%\end{equation}
The contractions of the elementary section with respect
to its maximal value $dl(\vec{v}_a)$
increase if the velocity of the section approaches
the boundary given by Eq. (\ref{eq:r76}).

\begin{figure}[htb]
\centerline{
\epsfig{file=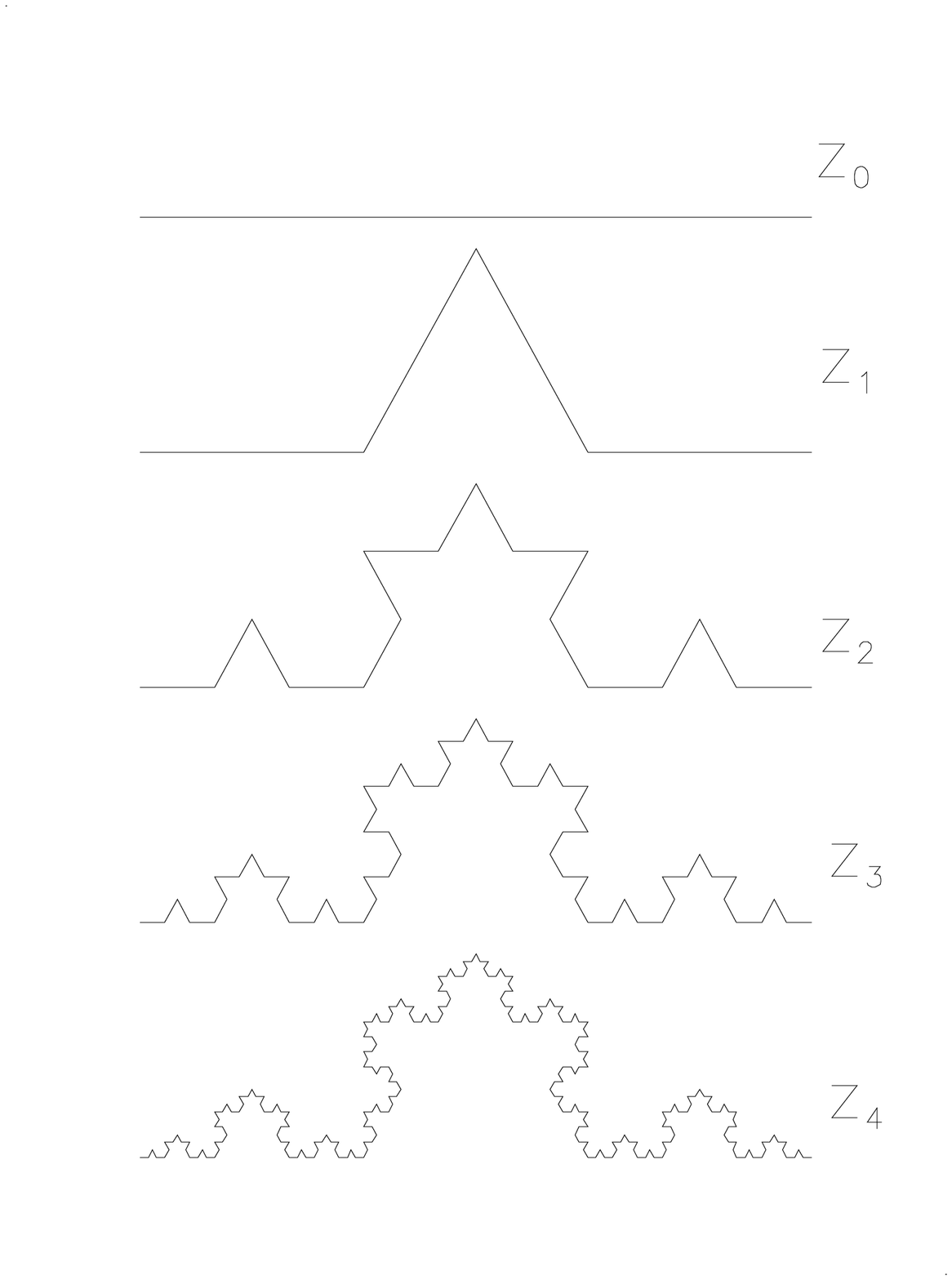,width=10cm,angle=0}
}
\caption{
Succesive approximations of the von Koch fractal curve. Its topological 
dimension is 1, while its fractal dimension is $\ln 4 / \ln 3$. 
}
\label{fig1PHOTS}
\end{figure}

\end{document}